\documentstyle[twoside,psfig,ijmpa1]{article}

\date{}
\begin{document}

\normalsize\textlineskip
\setcounter{page}{1}
\copyrightheading{}      

\vspace*{0.88truein}
\fpage{1}

\centerline{\bf SINGLE SPIN ASYMMETRIES IN INCLUSIVE HIGH ENERGY }
\vspace*{0.035truein}
\centerline{\bf HADRON-HADRON COLLISION PROCESSES}
\vspace*{0.37truein}
\centerline{\footnotesize LIANG ZUO-TANG} 
\vspace*{0.015truein}
\centerline{\footnotesize \it Department of Physics, Shandong University} 
\baselineskip=10pt
\centerline{\footnotesize \it Jinan, Shandong 250100, China}
\vspace*{10pt}
\centerline{\footnotesize C. BOROS}
\vspace*{0.015truein}
\centerline{\footnotesize \it 
      Department of Physics and Mathematical Physics,}
\centerline{\footnotesize \it 
                and Special Research Center for the
                Subatomic Structure of Matter,}
\centerline{\footnotesize \it 
                University of Adelaide,
                Adelaide 5005, Australia}
\baselineskip=10pt

\vspace*{0.21truein}
\abstracts{It has been realized for quite a long time 
that single-spin experiments, 
in which one of the colliding  
objects is transversely polarized, 
can be helpful in studying the 
properties of strong interaction in general and 
in testing Quantum Chromodynamics (QCD) in particular. 
Striking effects have been observed  
in the past few years which deviate drastically  
from the expectation of the perturbative QCD parton model. 
These effects have received much attention.
New experiments of the similar type are underway and/or planned. 
Different theoretical attempts have been made 
to understand these effects. 
In this review, the special role played by singly polarized 
high-energy hadron-hadron collisions 
in High Energy Spin Physics is emphasized. 
Characteristics of the 
available data for inclusive hadron productions 
are briefly summarized. 
Different theoretical approaches for such processes 
are reviewed with special attention to a non-perturbative 
model which explicitly 
takes the orbital motion of the valence quarks 
and hadronic surface effects into account. 
The connection between such asymmetries and 
hyperon polarization in unpolarized reactions 
is discussed.  
An example of the possible application of 
such experimental results in other processes is given.}{}{}

\vspace*{1pt}\textlineskip          
\section {Introduction}
\vspace*{-0.5pt}

\noindent         
Single-spin asymmetry study has recently received much
attention, both ex\-peri\-mentally$^{\ref{Cam85}-\ref{E70498}}$
and theoretically$^{\ref{Kane78}-\ref{Liang96}}$.
In fact, already in 1978, 
it was recognized\cite{Kane78} 
that such experiments, 
in which one of the colliding hadrons 
is transversely polarized, 
can be very useful in studying the properties of 
strong interaction in general and in testing 
Quantum Chromodynamics (QCD) in particular. 
These studies have been of particular interests 
in the Spin Community in recent years 
for the following reasons:\\[-0.66cm] 

\begin{itemize}
\itemsep=-0.10truecm
\item The experiments are conceptually very simple.
\item The observed effects are very striking.
\item Theoretical expectations based on the pQCD parton model 
      deviate drastically from the data. 
\item Information on transverse spin distribution and 
      that on its flavor dependence can be obtained from 
      such experiments.
\end{itemize} 
           
\vskip -0.26cm
A large amount of data is now available$^{\ref{Cam85}-\ref{E70498}}$.
Besides the well known 
analyzing power in $pp$-elastic scattering\cite{Cam85}, 
we have now data$^{\ref{Kle76}-\ref{E70498}}$ on
left-right asymmetries $A_N$ in 
single-spin inclusive processes  
for the production of various particles, such as 
different mesons, $\Lambda$ hyperons and direct photons, 
measured in experiments 
using various transversely polarized projectiles 
such as protons and antiprotons. 
Compared to elastic scattering, 
there are not only more data 
at higher energies 
but also more theoretical discussions. 
We will therefore focus on 
the inclusive processes in the following.

In single-spin single particle inclusive production experiments, 
one uses transversely polarized (or unpolarized) projectile 
to collide with unpolarized (or transversely polarized) target, 
and measure the inclusive cross section (or production rate) for
a given type of particle (or particle system) 
which enters the detector located, e.g., 
on the left-hand-side looking down stream. 
The asymmetry $A_N$ is defined as$^{\ref{Kle76}-\ref{E70498}}$,
\begin{equation}
A_N(x_F,p_\perp |h,s)\equiv 
     {N_L(x_F,p_\perp,h|s,\uparrow )-N_L(x_F,p_\perp,h|s,\downarrow )
\over N_L(x_F,p_\perp,h|s,\uparrow )+N_L(x_F,p_\perp,h|s,\downarrow )}.
\label{eq:Andef}
\end{equation}
Here, $h$ denotes the produced particle or particle system; 
$N_L(x_F,p_\perp,h |s,\uparrow )$ 
is the number-density of $h$'s 
which moves perpendicular to the polarization 
direction of incoming hadron and enters the detector 
when the colliding hadron is upwards polarized;  
$N_L(x_F,p_\perp,h |s,\downarrow )$
is the corresponding density 
for such $h$'s in the downwards polarized case; 
the subscript $L$ denotes that the detector is 
located on the left-hand-side looking down stream.  
$\sqrt{s}$ is the total center of mass (c.m.) energy of the 
colliding hadron system; $x_F\equiv 2p_\parallel /\sqrt{s}$, 
$p_\parallel $ is the longitudinal component of the momentum of $h$
in the c.m. frame and $p_\perp $ is the magnitude of the 
transverse component. 
Since $N_L(x_F,p_\perp;h|s,\downarrow )=N_R(x_F,p_\perp;h|s,\uparrow )$, 
the above mentioned definition of $A_N$ can also be written as,
\begin{equation}
A_N(x_F,p_\perp |h,s)\equiv 
     {N_L(x_F,p_\perp,h|s,\uparrow )-N_R(x_F,p_\perp,h|s,\uparrow )
\over N_L(x_F,p_\perp,h|s,\uparrow )+N_R(x_F,p_\perp,h|s,\uparrow )}.
\label{eq:Andef2}
\end{equation}
That is why the asymmetry is usually referred to as left-right asymmetry.
Obviously, the statistics would be very poor if one measured only those 
particles (or particle systems) which move perpendicularly to the 
polarization direction (strictly left or right). 
In practice, one makes use of particles produced 
in different transverse directions since 
the following is valid,
\begin{equation}
A_N(x_F,p_\perp |h,s)= {1\over \cos \varphi }
     {N(x_F,p_\perp,\varphi;h|s,\uparrow )-
      N(x_F,p_\perp,\pi-\varphi;h|s,\uparrow )
\over N(x_F,p_\perp,\varphi;h|s,\uparrow )+
      N(x_F,p_\perp,\pi-\varphi;h|s,\uparrow )}.
\label{eq:Andef3}
\end{equation}
where $\varphi$ is the azimuthal angle 
between the normal of the 
production plane and the polarization direction 
of the colliding hadron. 
Eq.(\ref{eq:Andef3}) is a direct consequence of space 
quantization for collision processes with spin-1/2 particles.

It should be mentioned that $A_N$ 
is the only parity-conserving asymmetry 
which can exist in single-spin single particle 
inclusive processes. 
This can be seen most clearly in the 
rest frame of the polarized colliding hadron. 
Here, the S-matrix can only be a function of 
the following three vectors: 
the polarization vector of the hadron $\vec S$, 
the incident momentum of the other colliding hadron $\vec p_{inc}$ 
and the momentum $\vec p$ of the observed particle or particle system.
Due to parity conservation, 
the S-matrix must be a scalar and the only spin-dependent 
scalar we can construct using these three vectors is 
$\vec S \cdot (\vec p_{inc} \times \vec p_h)$. 
We see that it is nonzero only 
if the transverse component of 
$\vec S$ is different from zero. 

It has been observed that $A_N$ 
is up to $40\%$ in the beam fragmentation region, 
whereas the theoretical expectation\cite{Kane78}
based on perturbative QCD parton model calculations 
were $A_N\approx 0$.
Different new theoretical approaches 
have been made in the last few years to understand 
such large discrepancies. 
New experiments are now underway and/or planned. 
It is therefore timely to summarize the 
presently available experimental results, 
the main ideas and results of different models 
in order to get a deep insight into the 
physics behind these phenomena and make 
full use of the wealth of the new experiments.  
Brief summaries in this direction 
have been made by Meng\cite{Meng95}
in the ``Workshop on the Prospects of Spin Physics at HERA'' 
held in August 1995 in Zeuthen, 
and by one of us\cite{Liang95,Liang96} 
in the ``Adriatico Research Conference on 
Trends in Collider Spin Physics'' held in December 1995 in Trieste, 
and in the ``XIII International Seminar on High Energy Physics
Problems: Relativistic Nuclear Physics  and Quantum Chromodynamics''
held in Dubna September 1996.  
The characteristics of the available data 
and main ideas and results of different models 
have been briefly summarized in these talks.
This is an extended version of those 
short summaries\cite{Meng95,Liang95,Liang96}.  
Part of it is also based on 
our doctoral theses\cite{Liang94,Boros96} at FU Berlin. 
The text in the following 
is arranged as follows: 
After this introduction, 
we will briefly summarize 
the characteristics of the existing data, 
the different approaches based on the  
pQCD  hard scattering model,   
the main ideas and results for $A_N$ of 
a non-perturbative approach of the Berliner group. 
They are given in section 2, 3 and 4 respectively. 
In section 5, 
we discuss the possibilities to differentiate these
different models by performing suitable further experiments. 
In section 6, we discuss the connection of $A_N$ to 
hyperon polarization in unpolarized reactions. 
An example for the possible applications of 
such striking experimental results 
is given in section 7.

\section {Characteristics of the data}

Experiments on single-spin inclusive hadron production 
was first carried out\cite{Kle76} in 1976 using the polarized proton beam at 
the Argonne Zero-Gradient Synchrotron (ZGS) to collide with 
unpolarized proton for $\pi^\pm$ production,
and later for $K$ and $\Lambda$ productions\cite{Drag78}.
Striking asymmetries have been observed, 
although the incident momentum of the beam was quite low 
(6 and 11.75 GeV/c). 
Subsequently, similar experiments 
were carried at CERN\cite{Ant80} and
at Brookhaven Alternating 
Gradient Synchrotron (AGS) 
at a bit higher energies and for the production of 
different  hadrons\cite{Bon88,Bon89,Sar90}.   
At Serpukhov, experiments 
was carried out\cite{Apo90}  
using 40 GeV/c pion-beam to collide with 
transversely polarized proton or deuteron targets.
More recently, a high energy (200GeV/c) 
transversely polarized proton and antiproton 
beam was obtained at Fermilab 
from the parity-violating decay of $\Lambda $ and $\bar \Lambda$. 
Using these beams, high energy single-spin experiments 
have been carried out$^{\ref{E70488}-\ref{E70496b}}$ 
for the production of different kinds of mesons and $\Lambda$. 
Very striking asymmetries have been 
observed$^{\ref{E70488}-\ref{E70496b}}$.  
Data are now available 
at $x_F\approx 0$ but rather high 
transverse momentum $p_\perp $ ($1\sim 4$ GeV/c) for 
$p(\uparrow )+p(0)\to \pi^0+X$  and 
at large $x_F$ but moderately large $p_\perp$
($0.2\sim 2$ GeV/c) for production of different kinds of 
hadrons and using proton and antiproton beams.
Here, as well as in the following of this review, 
we use the following notations: 
The first particle in a reaction denotes the projectile, 
the second denotes the target; $(\uparrow )$ means that the particle 
is transversely polarized, $(0)$ means that it is unpolarized.

Not only the $x_F$- but also 
the $u$-dependence have been studied in  
the lower energy experiments\cite{Kle76,Drag78}.  
[Here, as well as in the following, 
$s, t, u$ are the usual Mandelstam variables.]
Very striking features have been observed  
for the asymmetries at this energy. 
It has, in particular, been observed that 
the $x_F$-dependences for the asymmetries 
at different regions of $u$ 
are quite different from each other.
E.g., it has been observed that 
$A_N$ for $\pi^-$ is positive but small 
at $x_F\sim 0.5$ for both 
$u=0.2$(GeV/c)$^2$ and 
$u=-0.2$(GeV/c)$^2$ but 
it increases monotonically with increasing $x_F$ for 
$u=0.2$(GeV/c)$^2$ and reaches more than 30\% at $x_F\sim 0.8$ 
whereas decreases monotonically with increasing $x_F$ for 
$u=-0.2$(GeV/c)$^2$ and is even negative for $x_F>0.6$ 
and reaches -0.16 at $x_F\sim 0.8$. 
This is very interesting especially if 
we look at, in stead of $u$, the transverse momenta of the 
produced pion in the two cases, 
they are not very much different from each other:
those in case of $u=0.2$(GeV/c)$^2$ are around $0\sim 0.3$GeV/c 
and those in the case of $u=-0.2$(GeV/c)$^2$ are around 0.5GeV/c.
The existence of $A_N$ at these energies 
together with such striking features is 
still a puzzle for the theoretians. 
In fact, little theoretical discussion has yet 
been made in this connection. 
Whether the asymmetry in this energy region 
and that at the higher energies are of the same origin, 
whether the striking features observed here  
still exist in the higher energy experiments 
are questions which are still open.  
We recall that for small transverse momenta, that is
small positive $u$,  
diffractive dissociation may play an 
important role in particular at low energies. 
It is therefore also interesting and instructive 
to study these processes 
to see whether diffractive scattering play an important role and 
whether/how one can obtain useful information in connection with 
the understanding of the mechanism for 
diffractive scattering. 
This is an interesting topic which deserves 
much effort in the future.
In the following, we will concentrate on the higher energy 
region because almost all the theoretical discussions 
now available are in this region. 

To show the main features of these data 
at the high energy,  
we show in Figs. 1a, 1b, 1c and 1d the 
$A_N$'s for $p(\uparrow)+p(0)\to \pi+X$, 
$\bar p(\uparrow)+p(0)\to \pi+X$, 
$p(\uparrow)+p(0)\to \Lambda +X$ 
as functions of $x_F$  at moderate $p_\perp$, 
and in Fig. 1e  $A_N$'s for 
$p(\uparrow)+p(0)\to \pi^0+X$ and 
$\bar p(\uparrow)+p(0)\to \pi^0+X$ 
as functions of $p_\perp $ at $x_F\approx 0$. 
These data show the following characteristics:\\[-0.66cm] 

\begin{itemize}
\itemsep=-0.10truecm
\item[(1)] {\it $A_N$ is significant in, and only in, 
            the fragmentation region of the 
            polarized colliding object and 
            for moderate transverse momenta:}\\ 
            It can be seen that $A_N$ is approximately 
            equal to zero for $x_F\approx 0$ and different 
            $p_\perp $. For moderate $p_\perp $, 
            its magnitude increases monotonically 
            with $x_F$ and reaches, e.g., 40\% for 
            $p(\uparrow)+p(0)\to \pi^++X$ at $x_F\approx 0.8$.
\item[(2)] {\it $A_N$ depends on the flavor quantum number of the 
           produced hadrons:} \\
           It can be seen that $A_N$ in 
           $p(\uparrow )+p(0)\to \pi+X$ is positive for $\pi ^+$
           and $\pi^0$ but negative for $\pi^-$, and that 
           the magnitude of $A_N$ for $\pi^+$ and that for $\pi^-$ 
           are approximately equal but larger than that for $\pi^0$.   
\item[(3)] {\it $A_N$ depends also on the flavor 
           quantum number of the polarized projectile:}\\
           By using $\bar p(\uparrow )$ projectile instead of 
           $p(\uparrow )$, one observed that 
           while the asymmetry for $\pi ^0$ remains almost the same, 
           those for $\pi ^+$ and $\pi ^-$ exchange their roles.
\item[(4)] {\it $A_N\approx 0$ in the beam fragmentation
          region in $\pi ^- +p(\uparrow )\to \pi ^0 $ or $\eta +X$:}\\
          Measurement has also been made\cite {Apo90} 
          in the fragmentation of the pseudoscalar 
          meson $\pi^-$-beams (not shown in Figs.1a-e), 
          the results show that 
          in this region the asymmetry is consistent with zero.
\end{itemize} 

New experiments are underway and/or planned. 
Single-spin asymmetries for 
hadron and lepton-pair production will be carried 
out at RHIC\cite{RHIC} at $\sqrt{s}=200$GeV and at 
Serpukhov by RAMPEX Collaboration\cite{Serpukhov} 
at $p_{inc}=70$ GeV/c. 
Plans for similar experiments 
at HERA\cite{HERAN} have also been discussed. 
These experiments will certainly 
provided new insights into the origin 
of the observed large single-spin asymmetries.

\begin{center}
\begin{figure}
\psfig{file=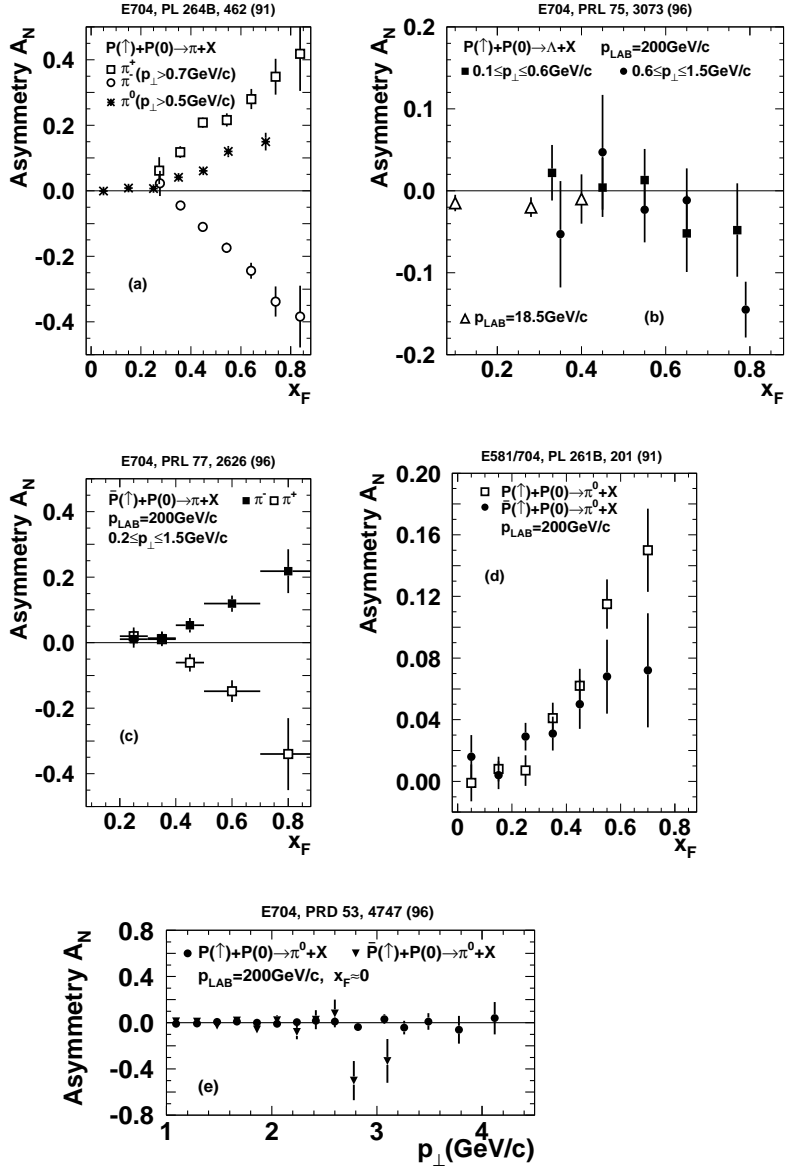,height=16cm}
\vskip 1.0truecm
\caption{Data for left-right asymmetries $A_N$'s from FNAL E704 
Collaboration.}
\end{figure}
\end{center}

\section{PQCD based hard scattering models}

Already in 1978, is was realized\cite{Kane78} that 
single-spin asymmetries can be very useful in studying 
the properties of hadron-hadron interactions in general and 
in testing the validity of pQCD calculations in particular. 
Using a straightforward 
generalization of the pQCD based hard scattering model 
to polarized case, Kane, Pumplin 
and Repko obtained\cite{Kane78} that, 
to the leading order in pQCD, 
$A_N\approx 0$ at high energies.  
This result is in clear disagreement 
with the data$^{\ref{Kle76}-\ref{E70498}}$ 
mentioned in the last section. 
Since pQCD calculations was very successful 
in describing unpolarized cross section 
even for transverse momentum of 
a few GeV (see, e.g. [\ref{pqcd}]), 
it was therefore a great surprise to see how large 
the discrepancy between the theoretical prediction\cite{Kane78}  
and the corresponding data$^{\ref{Kle76}-\ref{E70498}}$ is, 
and it is often referred as a 
challenge to the theoretians to understand this discrepancy. 

A number of mechanisms$^{\ref{Siv90}-\ref{BLM96}}$
have been proposed in last few years which can give 
rise to non-zero $A_N$'s
in the framework of QCD and
quark or quark-parton models.
As has been discussed in last section, 
the single-spin left-right 
asymmetries have been observed only 
for hadrons in the beam fragmentation regions and 
with moderate transverse momenta. 
As is well know, hadrons with large transverse momenta 
are products of processes with large momentum transfer and  
such processes are called ``hard processes''. 
For hard processes, impulse approximation 
can be used hence the constituents of the colliding 
hadrons can be treated as ``free particles'' and  
perturbative QCD calculations for the scattering amplitudes 
are valid.
It is also in such processes that 
the pQCD based hard scattering models 
were expected to, and indeed, work well\cite{pqcd}. 
In contrast, hadrons with small transverse momenta 
are dominated by the products of processes 
with  small momentum transfer. 
Such processes are called ``soft processes''\nobreak . 
For soft processes, perturbative QCD 
cannot be used and collective effects of 
the constituents in the colliding hadrons 
and other non-perturbative effects are important. 
Phenomenological models have to be used in describing these effects.
In studying the origin of the observed 
single-spin left-right asymmetries, 
we work in a kinematic region between the 
typical regions of the above mentioned two extreme cases. 
It is unclear whether perturbative methods are useful and 
whether typical ``soft effects'' play a role in this region.
This characteristics of the problem 
makes the investigation even more interesting  
since it is a suitable place to study the interplay between 
the ``soft'' and ``hard'' interactions 
in hadronic collision processes. 
It determines also that 
the theoretical approaches are
divided into the following two categories:
One of them starts from the ``hard'' aspects and try to 
include some of the influences from the ``soft'' aspects 
in some unknown factors.
The other starts from the ``soft'' side and neglects
most of the influences from the ``hard'' aspects. 
The former kind of approaches is usually referred  as 
perturbative QCD based hard scattering models 
and one of the well-known example of the latter is 
non-perturbative orbiting valence quark model.
The former have been discussed by many authors in 
literature$^{\ref{Kane78}-\ref{Ans95}}$, and
the latter is mainly pursued by the Berliner
group$^{\ref{Meng91}-\ref{BLM96}}$ and 
is sometimes referred as 
``Berliner Model'' or 
``Berliner Relativistic Quark Model (BRQM)'' 
for single-spin asymmetries. 
We review the main ideas and 
results of these two  approaches in the following. 
We start with pQCD based hard scattering models in this section and 
will deal with the second kind of models in the next section.

\subsection{PQCD based hard scattering 
model for high-$p_\perp$ hadron production 
in unpolarized hadron-hadron collisions}

\begin{center}
\begin{figure}
\psfig{file=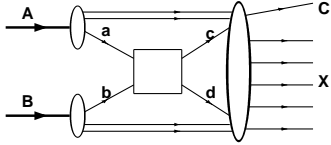,height=4cm}
\caption{Schematic illustration of factorization theorem in 
$A+B\to C+X$. The square in the center of the figure represents 
the elementary process $a+b\to c+d$ which can be calculated 
using perturbative QCD.}
\label{fig:fact}
\end{figure}
\end{center}

It is well known for a long time 
that in unpolarized hadron-hadron collisions, 
pQCD calculations can be applied for 
the production of high $p_\perp $ jets 
and/or high $p_\perp $ particles.
It has been shown that the 
inclusive cross section for hadron production 
can be factorized and thus be expressed as 
a convolution of the following three factors:
the momentum distribution functions of the constituents 
(quarks, antiquarks or gluons)
in the colliding hadrons;
the cross section for the elementary hard scattering
between the constituents of the two colliding hadrons; 
and the fragmentation function 
of the scattered constituent. 
E.g., for the inclusive process 
$A(0)+B(0)\to C+X$, this is illustrated 
in Fig. \ref{fig:fact}, and 
one has\cite{pqcd}, 

\vbox{
\[
E\frac {d\sigma}{d^3p} [A(0)B(0)\to C+X] 
=\sum_{abcd}\int dx_a dx_b dz_C 
\phi_{a/A}(x_a) \phi_{b/B}(x_b) \cdot 
\]
\begin{equation}
\phantom{XXXXXXXXXXXXXXXX}
{\hat{s}\over z_C^2\pi} {d\sigma \over d\hat{t}}(ab\to cd)
\delta(\hat{s}+\hat{t}+\hat{u}) D_F^{C/c}(z_C),
\label{eq:pqcdcsunpol}
\end{equation}
}

\noindent
Here, $p$ is the four momentum of $C$, whose longitudinal 
and transverse components are usually denoted by 
$p_\|=x_F\sqrt{s}/2$ and $\vec p_\perp$ respectively;  
$\phi_{a/A}(x_a)$ and $\phi_{b/B}(x_b)$ 
are the distribution function 
of the constituent $a$ in hadron $A$ and that of $b$ in $B$, 
where $x$ denotes the momentum fraction of hadron 
carried by the constituent; 
${d\sigma\over d\hat{t}}(ab\to cd)$ is the cross section for the 
elementary scattering process $ab\to cd$, 
where $\hat{s}, \hat{t}$ and $\hat {u}$ are the 
usual Mandelstam variables for the process; 
$D_F^{C/c}(z_C)$ 
is the fragmentation function describing the hadronization of $c$ into 
the hadron $C$ and anything else, where $z_C$ is the 
momentum fraction of the parton $c$ obtained by the hadron $C$.
While the distribution functions 
$\phi_{a/A}(x_a)$ and $\phi_{b/B}(x_b)$  
can be obtained by parameterizing the data from deeply 
inelastic lepton hadron scattering and other experiments 
and the fragmentation function $D_F^{C/c}(z_C)$ 
from phenomenological models,  
${d\sigma\over d\hat{t}}(ab\to cd)$ 
is the only factor which can be calculated using 
perturbation theory. 
From these calculations, one  
obtained not only the production rates 
but also the transverse momentum 
(or transverse energy) distributions 
of the high $p_\perp$ hadrons and/or jets. 

We note in particular here that, 
in the above mentioned simple version of 
the pQCD based hard scattering model, 
no intrinsic transverse momentum $\vec k_\perp $ of the constituents, 
$a$ and $b$, inside the hadrons, $A$ and $B$, is taken into account, 
and that the produced hadron in 
the fragmentation process is assumed to be in the 
same direction as the initial parton (i.e. $p=z_Cp_c$). 
We note also that\cite{pqcd} the model is very successful in describing 
the cross section for the production of hadrons or jets    
with large transverse momentum $p_\perp $ (say, $>5$GeV/c). 
The model can also be and has been 
applied to hadron or jet production 
in the region where $p_\perp $ is 
of the order of several GeV/c (say, 1 to 5 GeV/c), 
but in this region, effect of $k_\perp $-smearing,  
which is an effect due to the intrinsic transverse momentum 
of the constituents inside the hadron, 
is very significant and should be taken into account. 
This means that, to describe hadron production in this region, 
one should take a transverse momentum distribution 
for the constituents $a$ and $b$ in $A$ and $B$ into account, 
and make the following replacement in Eq.(\ref{eq:pqcdcsunpol}),
\begin{equation}
\int dx_a dx_b 
\phi_{a/A}(x_a) \phi_{b/B}(x_b) \longrightarrow 
\int dx_a dx_b d\vec k_{a\perp} d\vec k_{b\perp} 
\phi_{a/A}(x_a,\vec k_{a\perp} ) \phi_{b/B}(x_b,\vec k_{b\perp}).
\end{equation}
It is usually assumed\cite{pqcd} that 
the transverse distribution can be 
factorized from the longitudinal part, 
and a Gaussionian for $f(\vec k_\perp )$ 
was used to fit the data, i.e.,
\begin{equation}
\phi(x,\vec k_\perp )=\phi(x)f(\vec k_\perp ),
\end{equation}
\begin{equation}
f(\vec k_{\perp a})=
{e^{-k_\perp ^2/<k_\perp^2>}\over \pi  <k_\perp^2>}.
\end{equation}  
It has been obtained\cite{pqcd} that 
$<k_\perp^2>\sim 0.95$ (GeV/c)$^2$ which corresponds 
approximately to $<k_\perp>\sim 0.86$ GeV/c.

\subsection{Direct extension to polarized case}

The above mentioned pQCD based 
hard scattering models has been extended$^{\ref{Kane78}-\ref{Ans95}}$ 
in a straightforward manner to 
single-spin hadron-hadron collision processes. 
In this way, one obtains the following expression for 
the inclusive cross section for e.g. 
$A(\uparrow)+B(0)\to C+X$, 

\vbox{
\[
E\frac {d\sigma}{d^3p} [A(\uparrow)B(0)\to C+X] 
=\sum_{abcd,s_a,s_c}\int dx_a dx_b dz_c 
\phi_{a/A(\uparrow)}(x_a,s_a) \phi_{b/B}(x_b) \cdot
\]
\begin{equation}
\phantom{XXXXXXXXXXXXXX}
{\hat{s}\over z_C^2\pi} {d\sigma \over d\hat{t}}[a(s_a)b\to c(s_c)d]
\delta(\hat{s}+\hat{t}+\hat{u}) D_F^{C/c}(z_C;s_c),
\label{eq:pqcdcspol}
\end{equation}
}

\noindent
Here, $s_a$ and $s_c$ denote the spin of $a$ and that of $c$
respectively. 

Now, if we assume that the fragmentation function is independent of 
the spin of the scattered constituent $c$, i.e., 
\begin{equation}
D_F^{C/c}(z_C;-s_c)=D_F^{C/c}(z_C;s_c)\equiv D_F^{C/c}(z_C),
\label{eq:Df}
\end{equation}
we obtain the following expression for 
$E\Delta \frac {d\sigma}{d^3p} [A(tr)B(0)\to C+X]$ 
the difference between  
$E\frac {d\sigma}{d^3p} [A(\uparrow)B(0)\to C+X]$ and
$E\frac {d\sigma}{d^3p} [A(\downarrow)B(0)\to C+X]$ as 

\vbox{
\[
E \Delta \frac {d\sigma}{d^3p} [A(tr)B(0)\to C+X]=
\sum_{abcd} \int dx_a dx_b dz_C
\Delta \phi_{a/A(tr)}(x_a) \phi_{b/B}(x_b) \cdot 
\]
\begin{equation}
\phantom{XXXXXXXXXXXXXXXX}
{\hat{s}\over z_C^2\pi} 
\Delta {d\sigma \over d\hat{t}}[a(tr)b\to cd]
\delta(\hat{s}+\hat{t}+\hat{u}) D_F^{C/c}(z_C).
\label{eq:dcs1}
\end{equation}
}

\noindent
Here $\Delta \phi_{a/A(tr)}(x_a) \equiv 
\phi_{a/A(\uparrow )}(x_a,+) -
\phi_{a/A(\uparrow )}(x_a,-),$ where $+$ means 
$a$ is polarized in the same direction as $A(\uparrow)$, 
$-$ means in the opposite direction;  and,
\begin{equation}
\Delta {d\sigma \over d\hat{t}}[a(tr)b\to cd]\equiv 
{d\sigma \over d\hat{t}}[a(\uparrow)b\to cd]-
{d\sigma \over d\hat{t}}[a(\downarrow)b\to cd]
\label{eq:dM}
\end{equation} 
which is usually written as, 
\begin{equation}
\Delta {d\sigma \over d\hat{t}}[a(tr)b\to cd]=
 a_N[a(\uparrow) b\to cd]\cdot {d\sigma \over d\hat{t}}(ab\to cd).
\end{equation}
Here, $a_N$ is the single-spin asymmetry for the 
elementary process $ab\to cd$, 
\begin{equation}
 a_N[a(\uparrow)b\to cd]\equiv 
{\Delta {d\sigma \over d\hat{t}}[a(tr)b\to cd]\over  
{d\sigma \over d\hat{t}}(ab\to cd)}
\label{eq:aN}
\end{equation}
The cross section, or the scattering matrix ${\cal M}$, 
for the elementary hard scattering $ab\to cd$
can be calculated using pQCD. 
These  calculations are most conveniently 
performed using the helicity basis. 
Take $q_1q_2 \to q_1q_2$ as an example, 
and we obtain (See, e.g., [\ref{BSL80}]) 

\vbox{
\[
\Delta {d\sigma \over d\hat{t}}[q_1(\uparrow) q_2\to q_1q_2]\propto
2 Im \Bigl[ {\cal M}^*_{++,-+}({\cal M}_{++,++}+{\cal M}_{+-,+-})-
\phantom{XXXX}
\]
\begin{equation}
\phantom{XXXXXXXX}
{\cal M}^*_{++,+-}({\cal M}_{++,--}-{\cal M}_{+-,-+}) \Bigr],
\label{eq:ANqq}
\end{equation}
}

\noindent
where the subscripts of ${\cal M}$ denote the helicities 
of the particles in the initial and final states.
We see clearly from Eq.(\ref{eq:ANqq}) that
$a_N$ is nonzero only if the helicity-flip amplitude 
${\cal M}_{++,-+}$ is nonzero and has a phase difference 
with the helicity conserving amplitude(s) 
${\cal M}_{++,++}$ and/or ${\cal M}_{+-,+-}$, 
or the helicity flip amplitudes $\cal {M}_{++,+-}$ 
and $\cal {M}_{++,--}$ are nonzero 
and have a phase difference with each other
or $\cal {M}_{++,+-}$ and $\cal {M}_{+-,-+}$ are nonzero 
and have phase difference with each other. 
In either case, one needs at least one non-zero 
helicity flip amplitude to get a nonzero $\Delta {\cal M}$ 
thus a nonzero $a_N$.   
But, it is a well-known and remarkable property of 
perturbative QCD as well as perturbative QED that, 
in the limit of zero quark mass, helicity is conserved. 
That is, in this limit, all the helicity flip amplitudes 
vanish hence $a_N[q_1(\uparrow) q_2\to q_1q_2]=0$. 

Taking the quark mass $m_q$ into account, one obtains that,  
to the first order of pQCD, 
the helicity flip amplitude is nonzero but proportional to $m_q$. 
Using this, one gets\cite{Kane78}  
$a_N$ for the above mentioned 
QCD elementary process $q_1(\uparrow)+q_2(0)\to q_1+q_2$ 
is proportional to $m_q/\sqrt{s}$,
which is negligibly small at high energies.
Hence, we obtain that $A_N\approx 0$ 
at high energy $\sqrt{s}$.

The situation will not be changed much 
if we take an intrinsic transverse momentum 
distribution of the quarks in proton into account 
and assume that it is symmetric in the azimuthal direction.
Under such circumstances, we obtain 
$A_N\approx 0$ for hadron production.

\subsection{Comparison with data: 
            What conclusion can we draw now?}

We now compare the above mentioned result 
of the pQCD based hard scattering model 
with the available data$^{\ref{Kle76}-\ref{E70496b}}$. 
We are particularly interested in the following questions: 
Do the available data contradict QCD? 
Do the available data confirm QCD? 

The answer to the first question is: No! 
This is because  
pQCD works well only in the case 
in which the momentum transfer is large. 
Hence we expect that the above mentioned result is true only 
for the production of hadrons 
with high $p_\perp $ (say, $p_\perp \ge 5 $GeV/c). 
The only piece of data which is now available for high $p_\perp $ 
is that from E704 Collaboration for the production of $\pi^0$ 
and at $x_F\approx 0$. 
This data show that $A_N$ is indeed consistent with zero. 
(C.f. Fig.1e).

The answer to the second question is unfortunately also: No! 
This is because the only piece of data 
with which the prediction of pQCD can be compared is that 
for the production of $\pi^0$ and at $x_F\approx 0$.
There may be different reasons for the vanishing of 
$A_N$ in this case. 
E.g., it can be simply because 
the light sea quarks (antiquarks) 
which are responsible for the production 
of $\pi^0$ in this region 
are not polarized. 
It can be also because the polarizations of the 
$u$, $\bar u$, $d$, and $\bar d$ sea quarks (antiquarks) 
compensate with each other and therefore lead to zero $A_N$ 
for $\pi^0$ at $x_F\approx 0$. 
It does not necessary mean that the 
asymmetry from the elementary hard scattering 
process is zero, which is the prediction of QCD 
based on the perturbation theory. 
To test this prediction, one needs to measure $A_N$ for 
large $p_\perp $, large $x_F$ and at high energy. 
Large $p_\perp$ and high energy are necessary 
to guarantee the validity of 
the parton picture and that of the pQCD calculation, 
and large $x_F$ to guarantee 
that the quarks which contribute to the production of 
such hadrons are predominately the valence quarks 
of the polarized projectile and are transversely polarized 
before the interactions take place.
Such experiments can be carried out in the future. 
E.g., at RHIC\cite{RHIC}, the energy is already much 
higher than E704 experiment, hence,  
according to the above mentioned prediction 
of the pQCD parton model, 
the asymmetry $A_N$ at 
large $x_F$ high $p_\perp $ should be substantially smaller 
than that observed by E704.

One thing is however clear. 
That is, in the moderate $p_\perp $ and 
large $x_F$ regions, the above mentioned straightforward
extended version of pQCD hard 
scattering model contradicts the data. 
This means that improvements of the model
in this region are necessary. 

\subsection{New improvements}

The answer to the question of how to improve the model 
in order to describe the striking $A_N$'s observed 
in the moderate $p_\perp $ and large $x_F$ regions 
is in fact not  difficult to find. 
We recall that, in the model, 
the cross section is a convolution of three 
different factors, and the result $A_N\approx 0$ was obtained 
under the following approximations and/or assumptions:\\[-0.66cm]  

\begin{itemize}
\itemsep=-0.10truecm
\item[(i)] the left-right asymmetry $a_N$ 
for the elementary process was calculated 
using pQCD to the leading order; 
\item[(ii)] the distribution functions of 
the constituents in nucleon was assumed to be symmetric in 
intrinsic transverse motion;
\item[(iii)] the fragmentation function 
was assumed to be independent of the spin of the quarks.
\end{itemize}
\vskip -0.28cm
It is therefore also clear, to get a large asymmetry $A_N$, 
one can make use of one or more
of the following three possibilities:\\[-0.66cm]

\begin{itemize}
\itemsep=-0.10truecm
\item[(i)] Look for higher order 
 effects in the elementary processes
 which lead to larger asymmetries;
\item[(ii)] Introduce asymmetric
  intrinsic transverse momentum distributions
  for the transversely polarized quarks
  in a transversely polarized nucleon;
\item[(iii)] Introduce asymmetric transverse momentum distributions
  in the fragmentation functions for the transversely polarized quarks. 
\end{itemize}
\vskip -0.28cm
All these three possibilities have been discussed 
in the literature$^{\ref{Siv90}-\ref{Col94}}$.

It is clear that
under the condition that pQCD is indeed applicable for
the description of such processes,
it should (at least in principle) be possible to
find out how significantly the effects mentioned in (i)
contribute to $A_N$  by performing
the necessary calculations. 
In fact the calculations to the next leading order are 
not difficult to be carried out. 
The resulted asymmetry $a_N$ is still negligibly small 
at high energies.

In contrast to this,
the asymmetric momentum distributions 
mentioned in (ii) and (iii) can in principle be large.
But, these asymmetries 
have to be introduced by hand in the models. 
Whether such asymmetries indeed exist,
and how large they are if they exist,
are questions which
can only be answered by
performing suitable experiments.
In the following, we review the discussions of 
these different possibilities in more details.

\subsubsection{Asymmetric quark transverse momentum distribution?}

Possibility (ii) was first discussed by Sivers\cite{Siv90}. 
He argued that, to describe single transverse spin asymmetries, 
it is important to take the intrinsic transverse 
momentum $k_\perp$ of the quarks in nucleon into account
and to write the quark distribution functions 
as $\phi _{a/A}(x_a,\vec k_\perp)$. 
He assumed that, 
in a transversely polarized hadron $A(\uparrow)$, 
$\phi_{a/A(\uparrow)}(x_a,\vec k_\perp,s_a)$ can be asymmetric, i.e.
\begin{equation}
\phi_{a/A(\uparrow)}(x_a,-\vec k_\perp,\pm)\not=
\phi_{a/A(\uparrow)}(x_a,\vec k_\perp,\pm),
\label{eq:dqsiv}
\end{equation}
where $\pm$ means $a$ is polarized in the same ($+$) or the 
opposite ($-$) direction as the hadron $A$.
In this case, even if $a_N=0$, one obtains,

\vbox{
\[
E\Delta \frac {d\sigma}{d^3p} [A(\uparrow)B(0)\to C+X] 
=\sum_{abcd}\int dx_a dx_b d\vec k_\perp dz_C 
\Delta ^N \phi_{a/A(\uparrow)}(x_a,\vec k_\perp) \phi_{b/B}(x_b) \cdot
\]
\begin{equation}
\phantom{XXXXXXXXXXXXXX}
{\hat{s}\over z_C^2\pi} {d\sigma \over d\hat{t}}(ab\to cd)
\delta(\hat{s}+\hat{t}+\hat{u}) D_F^{C/c}(z_C),
\label{eq:dcssiv}
\end{equation}
}

\noindent
which can lead to $A_N$ that is significantly different from zero if 
\begin{equation}
\Delta ^N \phi_{a/A(\uparrow)}(x_a,\vec k_\perp)  \equiv
\sum_{s_a}[\phi_{a/A(\uparrow)}(x_a,-\vec k_\perp,s_a)-
\phi_{a/A(\uparrow)}(x_a,\vec k_\perp,s_a)],
\end{equation}
is significantly different from zero.

We recall that 
the effect of $k_\perp$-smearing plays an 
important role in describing the cross section of 
hadron production in unpolarized hadron-hadron collisions 
in the region of transverse momentum of several GeV/c, 
and that left-right asymmetry in single spin reactions 
have been observed$^{\ref{Kle76}-\ref{E70498}}$ 
mainly in the region of moderately 
large transverse momentum.
It is therefore quite natural and in fact even instructive 
to examine whether such effects due to 
the intrinsic transverse momenta of the quarks 
make significant contribution to these asymmetries. 
We note in particular that, 
the asymmetries have been observed$^{\ref{Kle76}-\ref{E70498}}$ 
for $p_\perp \sim 1$GeV/c and 
that the average intrinsic transverse momentum of the quarks  
has been determined\cite{pqcd} as $<k_\perp>\sim 0.86$ GeV/c. 
This implies that $k_\perp $ has to 
contribute to the $p_\perp $ significantly. 
Hence, it would be completely not surprising if it turns out that 
the $k_\perp $-distribution plays an important   
role in describing these asymmetry data$^{\ref{Kle76}-\ref{E70498}}$.
As mentioned in the last subsection, 
it can easily be shown that 
a symmetric intrinsic transverse momentum has 
little influence on the asymmetry. 
On the other hand, it is also clear that an asymmetric 
intrinsic transverse momentum distribution of the 
quarks in polarized hadron can manifest itself 
in the final state hadrons. 
It should lead to 
an asymmetric transverse momentum distribution for the 
produced hadrons if it indeed exists. 
It is also obvious that one can get a 
good fit to the data if one can choose the 
form and the magnitude of the 
asymmetric distribution freely.
It is therefore a natural and good question 
to ask whether the such asymmetric distribution 
indeed exists and if yes whether 
it is the major source for the asymmetries 
observed in the single-spin 
experiments$^{\ref{Kle76}-\ref{E70498}}$.

However, it has been shown by Collins\cite{Col93} that  
asymmetric intrinsic quark distribution 
cannot exist at leading twist, i.e. twist-2. 
At leading twist, the quark distribution is given by,
\begin{equation}
\phi_{a/A}(x,\vec k_\perp )\equiv 
\int {dy^-d^2\vec y_\perp \over (2\pi)^3} 
\exp(-ixP^+y^-+i\vec k_\perp \cdot \vec y_\perp) 
\langle P|\bar {\psi}_a(0,y^-,\vec y_\perp )
{\gamma^+\over 2}\psi_a(0)|P\rangle . 
\label{eq:phidef}
\end{equation} 
Here, $P$ is the four momentum of the hadron $A$; 
$P^+\equiv \frac{1}{\sqrt{2}}(P^0+P^3)$,
$\gamma^+ \equiv \frac{1}{\sqrt{2}}(\gamma^0+\gamma^3)$, 
and $y^-\equiv{1\over\sqrt{2}}(y^0-y^3)$ are 
light-cone variables.  
Since the transversely polarized state,  
e.g. $|\uparrow\rangle$, can be expressed in terms of 
the left- and right-handed 
helicity states $|L\rangle $ and $|R\rangle $, 
\begin{equation}
|\uparrow \rangle ={1\over \sqrt{2}}[|L\rangle + |R\rangle],
\end{equation} 
we have,
\begin{equation}
\phi_{a/A(\uparrow)}(x,\vec k_\perp)={1\over 2}[
\phi_{a/A}(x,\vec k_\perp;LL)+\phi_{a/A}(x,\vec k_\perp;RR)+
\phi_{a/A}(x,\vec k_\perp;LR)+\phi_{a/A}(x,\vec k_\perp;RL)].
\label{eq:phitran}
\end{equation}
Here, 
\begin{equation}
\phi_{a/A}(x,\vec k_\perp;h,h' )
\equiv \int {dy^-d^2\vec y_\perp \over (2\pi)^3} 
\exp(-ixP^+y^- +i\vec k_\perp \cdot \vec y_\perp) 
\langle P,h |\bar {\psi}_a(0,y^-,\vec y_\perp ){\gamma^+\over 2}
\psi_a(0)|P,h'\rangle ,
\label{eq:phidefpol}
\end{equation} 
where $h$ or $h'$ denotes 
the helicity ($L$ or $R$) of the hadron $A$.
It is clear that $\phi_{a/A}(x,\vec k_\perp;LL)$ and
$\phi_{a/A}(x,\vec k_\perp;RR)$ can only depend on the magnitude of 
$\vec k_\perp $ but {\it not} on the direction 
since no particular transverse direction 
is specified in this case. 
We obtain in particular that,
\begin{equation}
\phi_{a/A}(x,-\vec k_\perp;hh)=
\phi_{a/A}(x,\vec k_\perp;hh),
\label{eq:phil1}
\end{equation}
for $h =L$ or $R$.
Furthermore, using time reversal and parity inversion invariance, 
Collins\cite{Col93} obtained that
\begin{equation}
\phi_{a/A}(x,\vec k_\perp;LR)=\phi_{a/A}(x,\vec k_\perp;RL)=0.
\label{eq:phil2}
\end{equation}
It follows from Eqs.(\ref{eq:phitran}), (\ref{eq:phil1}) and 
(\ref{eq:phil2}) that,
\begin{equation}
\phi_{a/A(\uparrow)}(x,-\vec k_\perp)=
\phi_{a/A(\uparrow)}(x,\vec k_\perp)
\end{equation}
which contradicts Eq.(\ref{eq:dqsiv}).
This shows explicitly that 
time reversal and parity inversion invariances 
of strong interaction forbids 
the existence of such asymmetry at twist two. 
This is similar to that 
discussed by Christ and Lee more than 
twenty years ago\cite{Christ66} for 
deeply inelastic lepton-nucleon scattering, 
where these invariances 
forbid the existence of single-spin asymmetry. 

However, recently, 
Anselmino, Boglione, and Murgia\cite{Ans95} 
pointed out that 
the proof of Collins\cite{Col93} mentioned above 
is valid at and only at leading twist, 
where the initial state interactions between 
the constituents of the two colliding 
hadrons are neglected.
Asymmetric distributions 
are in principle allowed to exist 
at higher twists where the 
initial state interactions are taken into account. 
The suggestion of Anselmino {\it et al.} \cite{Ans95}  
is that these higher twist effects 
can be lumped into an effective 
asymmetric transverse momentum distribution 
in parton distribution functions.  
It is however unclear whether such asymmetric distribution  
indeed exists and, if yes, how large it is.
These are questions which cannot be answered in the model. 
The asymmetry has to be introduced 
by hand and can only be studied 
beyond the model or by experiments.
They therefore suggested\cite{HERAN} 
to measure these asymmetric distributions experimentally 
by measuring $A_N$ under the assumption that 
such asymmetric distributions are indeed 
responsible for the observed $A_N$. 
Obviously, such measurements make sense only in the case 
that one has tested such asymmetric distributions  
indeed exist and are responsible for 
the observed left-right asymmetries. 
The results would be useful if one has proved that they  
are universal in the sense that 
they can be used in different reactions. 
Since they are not intrinsic in the sense that they 
depend on the interaction of the hadron with the other, 
they cannot be measured in deeply inelastic lepton-nucleon scatterings 
using transversely polarized nucleon similar to that for 
unpolarized parton distributions.  
The ``measurement'' of the asymmetric 
quark distributions itself will provide unfortunately 
no deep insight into the problem. 
Furthermore, it is not  
clear whether factorization 
is valid at higher twists and 
whether expression (16) is a reasonable approximation 
to the full calculation. 
Therefore, one should analyze the scattering process by 
including higher twist contributions in the calculation  
right from the beginning.

\subsubsection{Higher twist parton distributions?} 

Higher twist effects in connection with single-spin  
asymmetries were discussed first by  
Efremov and Teryaev\cite{Efr82,Efr95} 
and later by Qiu and Sterman\cite{Qiu91,Qiu98}.   
It has been argued\cite{Qiu98} that, 
the results of the calculations by 
Kane, Pumplin and Repko\cite{Kane78}, 
which show that $A_N$ is proportional to the quark mass $m_q$, 
indicate already that $A_N$ is a higher twist effect. 
Qiu and Sterman have therefore analyzed\cite{Qiu98} the possible 
contributions of the next to leading twist terms, i.e. 
the twist-3 contributions.   
Since the asymmetries are observed mainly in the fragmentation region, 
where contributions of the valence quarks of the projectile dominate, 
Qiu and Sterman used a ``valence-quark-soft-gluon'' approximation where 
only interaction of valence-quarks from one colliding hadron with 
gluons from the other are taken into account. 
In this way they simplified the problem in a great deal. 
They argued that the leading contributions should come from 
the term containing the following twist-3 parton distribution,
\begin{equation}
T^{(V)}_{Fa}(x_1,x_2)
=\int {dy_1^-dy_2^- \over 4\pi} 
 e^{ix_1P^+y_1^-+i(x_2-x_1)P^+y_2^-}
\langle P,\vec s_\perp|\psi_a(0)\gamma^+
[\varepsilon ^{\rho s_\perp n\bar n}F_\sigma^+(y_2^-)]
\psi_a(y_1^-)|P,\vec s_\perp\rangle 
\label{eq:Tvdef}  
\end{equation}
where $\varepsilon^{\rho s_\perp n\bar n}\equiv 
\varepsilon^{\rho\sigma\mu\nu}\vec s_{\perp \sigma}n_\mu\bar n_\nu$, 
$n\equiv (n^+,n^-,\vec n_\perp)=(0,1,\vec 0_\perp )$, 
$\bar{n}\equiv (\bar{n}^+,\bar{n}^-,\vec{\bar{n}}_\perp)=(1,0,\vec 0_\perp )$. 
They showed that the existence of such twist-3 parton distribution and 
the exchanged terms of the leading and next-to-leading graphs can
lead to an significant left-right asymmetry for the produced particle.  

Compared with the definition of the twist-2 parton distributions 
in Eq.(\ref{eq:phidef}), 
we see that while there are two field operators in Eq.(\ref{eq:phidef}) 
which can thus be interpreted as probability distribution, 
there are three filed operators in Eq.(\ref{eq:Tvdef}).
One cannot interpret $T^{(V)}_{Fa}(x_1,x_2)$ as probability distribution. 
Since it depends on two fractional momentum $x_1$ and $x_2$, it 
is called correlation function.
  
It is usually expected that a twist-3 contribution to $A_N$ 
should be proportional to $\mu /p_\perp$ and thus decreases 
with increasing $p_\perp $ (where $\mu $ is a non-perturbative scale for 
the twist-3 matrix element).
However, based on the above mentioned approximation, 
Qiu and Sterman showed that the $\mu/p_\perp $ term 
is not the dominating term in the twist-3 contribution to the $A_N$ 
in the E704 kinematic region. 
From dimensional analysis alone, one would expect that there 
are two types of twist-3 contributions, 
one of which is proportional to 
$\mu p_\perp /(-u)$ 
and the other is proportional to 
$p_\perp \mu/(-t)$. 
We note that, for high energy $\sqrt{s}$, moderately large $p_\perp $ and
large $x_F$, $-t\sim p_\perp ^2$, $-u\sim s$. 
Thus the two terms are proportional to 
$\sim \mu p_\perp /s$ and $\sim \mu/p_\perp $ respectively. 
We see that while the former vanishes at $s\to \infty$, 
the latter is suppressed at high $p_\perp$. 
For not very large energy $\sqrt{s}$, 
the former, which is proportional to $p_\perp $, 
can be larger than the latter in particular 
in the large $p_\perp $ region.
The results of Qiu and Sterman showed that  
at the FNAL E704 energy, the first type of contribution 
dominates at high $x_F$ moderate $p_\perp $
and that $A_N$ changes rather smoothly with increasing $p_\perp$.
They found also that $A_N$ is mainly proportional to 
$T^{(V)}_{Fa}(x_1,x_2=x_1)$.

A simple model for $T^{(V)}_{Fa}(x,x)$ was proposed\cite{Qiu98}. 
Since, for $x_1=x_2=x$, one has,
\begin{equation}
T^{(V)}_{Fa}(x,x)
=\int {dy_1^-dy_2^- \over 4\pi} 
 e^{ixP^+y_1^-}
\langle P,\vec s_\perp|\psi_a(0)\gamma^+
[\varepsilon ^{\rho s_\perp n\bar n}F_\sigma^+(y_2^-)]
\psi_a(y_1^-)|P,\vec s_\perp\rangle ,   
\end{equation}
which is similar to the twist-2 parton distribution 
[Eq.(\ref{eq:phidef})] with an extra factor 
$[\varepsilon ^{\rho s_\perp n\bar n}F_\sigma^+(y_2^-)]$.
Qiu and Sterman argued\cite{Qiu98} 
this extra factor leads to no extra $x$ dependence, 
thus they assume,
\begin{equation}
T^{(V)}_{Fa}(x,x)=\kappa _a\lambda q_a(x),
\end{equation}
where $\kappa _a$ denotes the sign of $T^{(V)}_{Fa}$ and 
$\lambda$ is a constant describing its magnitude. 
From the signs of data$^{\ref{E70488}-\ref{E70498}}$ 
of $A_N$ for $\pi^+$ and $\pi^-$, they determined that 
$\kappa _u=1$ and $\kappa_d=-1$. 
The magnitude of $A_N$ can also  
be roughly reproduced by
taking $\lambda =0.080$GeV.  
By using this simple model for $T_{Fa}^V(x,x)$, 
they demonstrated also how the twist-3 
contribution to $A_N$ increases with increasing $x_F$.  

This result is rather interesting and 
attractive for the following reasons:
First, it directly extends the 
pQCD based hard scattering model to twist-3.  
If it turns out to work well, 
this opens another field to test 
the applicability of pQCD.
Second, the results show that twist-3 effects 
at the energy region 
like that of the FNAL E704 experiments do not vanish at high $p_\perp $.
This provides us a good opportunity to study twist-3 distributions, which 
have poorly studied yet.

On the hand, this result is still somewhat unsatisfactory 
in particular from the phenomenological point of view, 
since it does not supply a physical picture for 
the single spin asymmetries. 
The correlation functions, which 
characterize the soft-hadron properties 
responsible for the asymmetries, 
are to be extracted from the experiments 
rather than to be given by the  theory.    
The two parameters, $\lambda$ and $\kappa_a$, 
which determine respectively 
the sign and the magnitude of $T_{Fa}^{(V)}$, 
are free parameters in the model which 
are determined by the $A_N$ data. 
It is therefore unclear whether they are indeed 
the major source for the observed $A_N$.
Furthermore, there seems to be an inconsistency in the 
restriction to twist three, at least 
as far as the understanding of the present data 
is concerned.   
We can easily see that, if we can obtain from the 
twist-3 contribution an $A_N$ as large as $30\%$ to $40\%$, 
the twist-3 contribution to the cross section should at least 
half or $60\%$ as large as the leading twist-2 contribution. 
This is then not a small correction any more.
How about the twist-4 and even higher twist contributions?  
According to the above mentioned dimensional analysis, 
do they vanish or even increase as 
$p_\perp ^2$ with increasing $p_\perp$?  
While we expect  
pQCD calculations with higher twist effects included   
to work for larger $p_\perp$ values,   
it seems that a modeling of the soft-hadronic 
properties is necessary to understand the physical 
origin of (the present) data.  

\subsubsection{Spin dependent fragmentation function?}

Since fragmentation of quark is a basically soft process 
containing contribution from higher twists,
it is conceivable that fragmentation effects could account 
for at least  part of the single-spin asymmetries.  

After having shown that the quark intrinsic transverse 
momentum cannot be asymmetric,  
Collins, Heppelmann and Ladinsky
have argued that\cite{Col93,Col94} the origin for 
the observed left-right asymmetry can only 
be the fragmentation, i.e. possibility (iii) mentioned above 
should be true. 
They suggested that transverse momentum for hadron obtained in 
the fragmentation process with respect to (w.r.t.) 
the initial parton 
is important in the description of single-spin asymmetries 
and that the fragmentation function 
depends on the spin of the quark. 
It can be asymmetric in transverse momentum distribution 
if the quark is transversely polarized. 
More precisely, one should replace 
the $D_F^{C/c}(z_C;s_c)$ 
in Eq.(\ref{eq:pqcdcspol}) by 
\begin{equation}
D_F^{C/c}(z_C,\vec k_{F\perp};s_c)= 
D_F^{C/c}(z_C,\vec k_{F\perp})(1+ \alpha \vec s_c \cdot 
{\vec p_c\times \vec k_{F\perp} \over |\vec p_c\times \vec k_{F\perp}|}),
\label{eq:frag}
\end{equation}
where $\vec k_{F\perp}$ is the transverse momentum 
of $C$ w.r.t. the momentum of the parton $c$.
Based on this assumption, 
the authors of [\ref{Art97}] 
have constructed a simple model based on the string fragmentation 
model\cite{And83} and have 
calculated $A_N$ for meson production.

We compare this possibility with 
possibility (ii) mentioned in the last subsection, 
and we see the following. 
First, just as the assumption of the 
existence of the asymmetric 
quark transverse momentum distribution, 
the existence of spin dependence of the fragmentation function 
cannot be derived from the model, 
it has to be introduced by hand.
Whether it indeed exists and, if yes, how large it is, 
has to be studied beyond the model and/or by experiments. 
Second, compared with the possibility (ii), 
it has however a great advantage that 
this assumption can be tested directly by 
performing suitable experiments!
It is possible to have processes where 
only fragmentation effects play a role.   
Studying such processes can provide direct information 
on the spin effects in fragmentation processes.
Presently, there exist already a number different  
measurements which directly or indirectly 
suggest that this effect cannot be large. 
In this connection, we have the following.

(a) That the fragmentation function can be dependent of 
the spin of the quark was first discussed 
for longitudinally polarized case 
in 1977 by Nachmann\cite{Nac77}, 
in 1978 by Efremov\cite{Efr78} 
and more recently by 
Efremov, Mankiewicz, Tornqvist\cite{Efr92}.  
These authors argued that the fragmentation can be dependent of 
the spin of the fragmented quark and 
introduced a quantity which they called ``handedness'' 
of the jet produced by the fragmentation of the quark. 
They suggested that the jet handedness 
should be significantly different from zero if 
quark fragmentation is indeed dependent of 
the helicity of the quark. 

We recall that, jet handedness $H$ 
is defined as$^{\ref{Nac77}-\ref{SLD95}}$,
\begin{equation}
H\equiv \frac{N_{\Omega<0}-N_{\Omega>0}}{N_{\Omega<0}+N_{\Omega>0}},
\end{equation}
where $\Omega\equiv \vec e \cdot (\vec p_1\times \vec p_2)$, 
$\vec e$ is the unit vector along the jet axis, 
$\vec p_1$ and $\vec p_2$ 
are the momenta of two particles in the jet 
chosen in a charge independent way such as 
$|\vec p_1|>|\vec p_2|$.  
This quantity $H$ has been measured recently by 
SLD Collaboration at SLAC\cite{SLD95}. 
They found an upper limit of 0.063 for $H$ at 95\% c.l.
This result shows that the 
dependence of fragmentation 
on the spin of the quark 
in the longitudinally polarized case is very weak.
If there is such a dependence at all, 
the corresponding asymmetry that one 
obtains in the transversely polarized case 
should be less than 10\%, which is far from enough 
to account for the left-right asymmetries observed 
in single-spin hadron-hadron collisions$^{\ref{Kle76}-\ref{E70498}}$.

(b) The suggestion\cite{Col94} that the fragmentation results 
depend on the spin of the initial quark 
in the transversely polarized case has in fact 
a model realization i.e. the LUND model discussed 
by Andersson, Gustafson and Ingelman\cite{And79} in 1979 
in connection with hyperon polarization\cite{Hel96} 
in unpolarized hadron-hadron collisions.  
Here, a semi-classical picture was proposed to explain 
the striking $\Lambda$ polarization observed in unpolarized 
hadron-hadron or hadron-nucleus collisions.  
We note that hyperon polarizations in unpolarized 
hadron-hadron collisions have been observed mainly in 
the fragmentation regions of the colliding hadrons 
and that $\Lambda$'s in fragmentation regions 
in $pp$ or $pA$ collisions are predominately 
fragmentation products of spin-zero $ud$-diquarks. 
In Lund model, it was argued that, 
if a quark-antiquark pair $s\bar s$ is 
produced in the fragmentation of a spin zero $ud$-diquark 
with the $s$-quark has a transverse momentum to the left, 
there has to be a space separation between 
the $s$ and $\bar s$ since a piece of string is needed to  
generate the mass and transverse energy of the $s\bar s$.
Hence, the $s$ and $\bar s$ should have a 
relative orbital angular momentum 
pointing upwards. 
According to angular momentum conservation, 
the spins of the $s$ and $\bar s$
should have a large probability to point downwards 
to compensate this relative orbital angular momentum. 
One expects therefore to see a downwards polarized 
$\Lambda$ since the spin of $\Lambda$ 
is completely determined by its $s$-quark.
This qualitative result is indeed 
in agreement with the data\cite{Hel96} 
for $\Lambda$ production in $pp$ or  
or $pA$ collisions. 
 
We now apply the arguments to fragmentation 
of an upwards polarized quark. 
We note that, 
to produce a pseudoscalar meson such as a pion 
with a moderately large transverse momentum, 
one needs an antiquark which is downwards polarized. 
The transverse momentum of this antiquark 
should therefore have a large probability 
to the left so that the relative angular momentum between 
the produced $\bar q$ and $q$ can compensate their spins. 
This leads to an asymmetric transverse momentum distribution 
for the produced hadrons.  
We see that such a simple picture 
indeed leads to a left-right asymmetry 
for the hadrons produced in the fragmentation 
of a transversely polarized quark. 

Calculations based on LUND model for $\Lambda$ polarization 
in unpolarized $pp$ collisions have been made\cite{And79}. 
The results are substantially smaller than data\cite{Hel96}.
Furthermore, if this is indeed the origin of 
the $\Lambda$ polarization observed in hadron-hadron collisions, 
one would expect no $\Lambda$ polarization in 
the beam fragmentation region of 
$K^-+p(0)\to \Lambda +X$ 
where $s\to \Lambda +X$ gives the dominate contribution. 
This is in contradiction with the data\cite{Gou86} 
which shows that $\Lambda$'s here 
are positively polarized and 
the magnitude of the polarization is quite large. 
Furthermore, measurements of 
$\Lambda$ transverse polarization in $e^+e^-$
annihilation have also been carried out 
by TASSO at DESY\cite{Tasso85} 
and ALEPH at CERN\cite{Aleph96}. 
Here, only fragmentation effects contribute.
The results\cite{Tasso85,Aleph96} 
of both groups show that transverse $\Lambda$ polarization 
is consistent with zero in $e^+e^-$ annihilation into hadrons. 

All these different experiments suggest that 
spin dependence of fragmentation function is, 
if at all, quite weak and can not be the major source of 
the left-right asymmetries observed$^{\ref{Kle76}-\ref{E70498}}$ 
in high energy singly polarized hadron-hadron collisions. 
It would also be difficult to 
understand why $A_N$ for 
$p(\uparrow\nobreak)+p(0)\to \Lambda +X$ 
is\cite{E70496b} quite large in magnitude for large $x_F$, 
if fragmentation is the major source for the 
existence of such asymmetries. 
This is because here fragmentation of 
spin zero $(u_vd_v)_{0,0}$ diquarks dominates. 
Since this is a spin zero object, 
there should be no left-right asymmetry for 
the produced $\Lambda$  
w.r.t. the moving direction of this $ud$-diquark. 
Hence, if fragmentation is the major source of $A_N$, 
one would expect that $A_N$ for $\Lambda$ in 
the very large $x_F$ region is very small, 
which contradicts the E704 data\cite{E70496b}. 
Further experimental tests can 
and will be carried out. 
Experiments by HERMES at HERA will be able 
test this assumption directly. 
Further suggestions to test this 
can also be found in section 5 below.

\section {A non-perturbative approach}

As has been seen in the last section, 
the ``pQCD based hard scattering models'' start from 
the ``hard'' side and try to include 
the influences from the ``soft'' 
aspects in some of the unknown factors. 
The model itself cannot determine whether these ``soft'' effects 
indeed exist, and if yes, how large they are. 
It is also not clear which effect plays the most important role 
in describing single-spin asymmetries. 
It is therefore useful and necessary to have phenomenological 
model studies in order to get deeper insight into the physics 
behind these possible effects and to find out which effect 
plays the dominating role. 
Several approaches$^{\ref{Tro95}-\ref{BLM96}}$ 
of this kind have been proposed recently.
These approaches start directly 
from the ``soft'' side and thus offers the possibility 
to study the origin of the ``soft effects'' in 
an explicit manner. 
The most successful one is perhaps     
the ``orbiting valence quark model'' proposed by 
the Berliner group$^{\ref{LM92}-\ref{Boros96}}$.
This model is sometimes also referred as ``Berliner Model'' 
or ``Berliner Relativistic Quark Model (BRQM)'' 
for single-spin asymmetries.  
In this section, we will concentrate 
ourselves on this approach.
The materials we presented here 
are based on the publications$^{\ref{LM92}-\ref{Boros96}}$  
of the group in this connection 
and the two doctoral theses\cite{Liang94,Boros96}  
from us at FU Berlin. 
 
\subsection{Orbiting valence quarks in polarized nucleon}

The existence of orbital motion of quarks 
in nucleon have been studied by many author in different 
connections$^{\ref{Seh74}-\ref{LR97},\ref{LM92}-\ref{BLM96}}$.
We note that, 
various experimental facts show that 
valence quarks in a light hadron should be treated as Dirac particles 
moving in a confining field created by other constituents 
of the same hadron.  
The masses of these valence quarks are much smaller 
than the proton mass. 
This is very much different from 
what one expected\cite{Mor65} in 1960s 
that the masses of the quarks 
in nucleon should be of the order of several GeV and is larger 
than the proton mass so that non-relativistic approximation 
can be used in describing the motion of these quarks 
in the nucleon. 
In contrast, it is now a well established fact that 
the masses of the valence quarks of the nucleon are much 
smaller than the proton mass and is also negligbly small 
compared to their kinetic energies in nucleon. 
Hence, to describe their motion inside a nucleon, 
it is the ultra-relativistic limit rather than the 
non-relativistic approximation should be used.    
It is known that Dirac particles confined in a limited space have a
number of remarkable properties,  and one of these features 
(which is almost trivial but very important) is the following: 
Independent of the details of the
confining system, the orbital angular momentum of this particle is {\it not} a
good quantum number --- except in the non-relativistic limit.                   
That is, except for those cases in which the ratio between the kinetic energy
and the mass of the quark  is much less than unity, 
the eigenstates of the quarks {\it cannot} be characterized 
by their orbital angular momentum quantum numbers.                                       This implies that one can never simply say that 
the valence quark has a definite orbital angular momentum $l=0$. 
Hence, orbital motion is always involved for  
the valence quarks even when they are in their ground states. 

A demonstrating example,  
in which the confining potential is taken as central, 
is given in [\ref{LM92}] and [\ref{Liang94}].  
In this case, stationary states should be characterized by 
the following set of quantum numbers 
$\varepsilon,j,m$ and ${\cal P}$ 
which are respectively the eigenvalues of the
operators $\hat H$ (the Hamiltonian), 
${\hat {\vec j}}^2, \hat j_z$ (the total angular
momentum and its $z$-component) and $\hat {{\cal P}}$ (the parity), and the 
ground state, which is characterized by 
$\varepsilon=\varepsilon_0,$
$j=1/2 ,m=\pm1/2$ and ${\cal P}=+$, 
is given by, 
\begin{equation}
\psi_{\varepsilon _0{1 \over 2}m+}(r,\theta,\phi) =
 \left( \matrix{& f_{00}(r)\ \Omega^{{1 \over 2}m}_0  (\theta,\phi) \cr\cr
 &g_{01}(r)\ \Omega^{{1 \over 2}m}_1 (\theta, \phi) \cr}\right),
\label{eq:qwav} 
\end{equation}
where,
{\small 
\begin{equation} 
\Omega^{{1 \over 2}m}_0(\theta,\phi)= 
Y_{0 0} (\theta,\phi)\ \xi(m), 
\end{equation} 
\begin{equation} 
\Omega^{{1 \over 2}m}_1(\theta,\phi) = 
- \sqrt{{3-2m \over 6}} Y_{1\ m-{1\over 2}}(\theta,\phi)\ \xi({1\over 2})+ 
  \sqrt{{3+2m \over 6}} Y_{1\ m+{1\over 2}}(\theta,\phi)\ \xi(-{1\over 2}). 
\end{equation} }
Here, $\xi (\pm {1 \over 2})$ 
stand for the eigenfunctions for the spin-operator
  $\hat\sigma_z$ with  eigenvalues $\pm 1$, and 
  $Y_{\ell \ \ell_z}(\theta,\phi)$ for the 
  spherical harmonics which form a standard 
  basis for the orbital angular
  momentum operators $({\hat {\vec \ell}}^2, \hat {\ell}_z)$.
The radial part, $f_{00}(r)$ and $g_{01}(r)$, 
of the two-spinors are determined by the Dirac equation 
for given potentials.   
Taking the $j_z\equiv m=+{1 \over 2}$ state as example, 
we obtain,
\begin{equation}
\langle \hat l_z (\varepsilon _0,{1 \over 2},{1 \over 2},+)\rangle 
={2\over 3} \int _0^{\infty} g_{01}^2(r) r^2 dr > 0, 
\label{eq:lz}
\end{equation}
\begin{equation}
\langle \hat l^2 (\varepsilon _0,{1 \over 2},{1 \over 2},+)\rangle 
=2\int _0^{\infty} g_{01}^2(r) r^2 dr > 0.
\label{eq:l2}
\end{equation}
Also, the current
density $J^\mu= (\rho,\vec J)=\bar\psi \gamma^\mu\psi$ is given by,
\begin{equation}
\rho(\varepsilon_0,{1 \over 2},{1 \over 2},+|\vec r) 
={1 \over 4\pi} [ f_{00}^2 (r) + g_{01}^2(r)],
\end{equation}
\begin{equation}
J_x(\varepsilon_0,{1 \over 2},{1 \over 2},+|\vec r) 
= + { y\over 2\pi r} f_{00}(r)g_{01}(r), 
\label{eq:Jx}
\end{equation}
\begin{equation}
J_y(\varepsilon_0,{1 \over 2},{1 \over 2},+|\vec r) 
= - { x\over 2\pi r} f_{00}(r)g_{01}(r), 
\label{eq:Jy}
\end{equation}
\begin{equation}
J_z(\varepsilon_0,{1 \over 2},{1 \over 2},+|\vec r)= 0.  
\phantom{{ x\over 2\pi r} f_{00}(r)g_{01}(r)} 
\label{eq:Jz}
\end{equation}
Here, we explicitly see that, even in the ground state,
$\langle \hat l_z \rangle \not = 0$,
$\vec J(\vec r)\not = 0$.
This implies that orbital motion is always involved for 
the valence quarks in the nucleon. 
We see also that $\langle l_z\rangle >0$ implies 
that the momentum distributions of these valence quarks 
and their spatial coordinates 
inside the hadron must be correlated in a given manner.  
This can be seen by examining at the $p_y$-distribution 
of this valence quark at different $x$, 
namely in different plane parallel to the $oyz$-plane.
It has been demonstrated that\cite{Liang94} 
for $x>0$, $p_y$ has an extra 
component in the positive $y$-direction i.e. 
$\langle p_y \rangle >0$;
but for $x<0$, $p_y$ has an extra 
component in the negative $y$-direction i.e.
$\langle p_y \rangle <0$. 

These features of the intrinsic motion of the quarks 
inside nucleon, in particular the distribution 
of transverse momentum mentioned above,  
are very interesting and they should be 
able to manifest themselves in hadron production 
in hadron-hadron collisions.   
To study this, we need to know how the quarks 
are polarized in a polarized nucleon. 
This is determined by the wavefunction of the nucleon. 
It has also been shown\cite{LM92,Liang94} 
that the wave function of the proton 
in such a relativistic quark model can be obtained 
simply by replacing the Pauli spinors in the static quark models 
by the corresponding Dirac spinors. 
Both of them are direct consequence of Pauli principle and 
the assumption that baryon is in the 
color singlet state hence the color degree is 
completely antisymmetric.

The wave function has two direct consequences:
First, it can be used to calculate 
the baryons' magnetic moments $\mu_B$ in terms 
of the magnetic moments $\mu_u$, $\mu_d$ and $\mu_s$  
of the valence quarks. 
An explicit expression of $\mu_B$ in terms of  
$\mu_u$, $\mu_d$ and $\mu_s$  
has been obtained in [\ref{LM92}]. 
The results show that this expression
is exactly the same as that 
in the static quark model 
and it is independent 
of the confining potentials. 
The only difference is that 
$\mu_u$, $\mu_d$ and $\mu_s$ are 
free parameters in the static models, 
they are constants depending on the 
choices of the confining potentials 
in the relativistic models.
This is very interesting since it 
shows that such relativistic models are 
as good (or as bad) as the static models in 
describing the baryons' magnetic moments data.
It explains also why the static models give 
a reasonable description of baryons' magnetic moments 
although we know that the quark masses are not as 
large as one thought in the 1960s to justify 
the use of non-relativistic limit. 
It explains also why the obtained values for 
$\mu_u$, $\mu_d$ and $\mu_s$ in the static models 
by fitting the data are not simply $1/(2m_q)$. 
Second, the polarization of the valence quarks 
is also determined 
by the wave function of the nucleon. 
This implies that, for proton, 
$5/3$ of the 2 $u$ valence quarks are polarized in the same,  
and $1/3$ in the opposite, direction as the proton. 
For $d$, they are $1/3$ and $2/3$ respectively.    
We see that both of them are polarized and the polarization is flavor
dependent.

\subsection{Production mechanism for hadrons in the fragmentation region 
            in hadron-hadron collisions}

In order to study the asymmetries observed in 
single-spin reactions, we need to know the production 
mechanism of the hadron in such reactions.
Since the asymmetries are observed mainly 
in the fragmentation region, 
we will also concentrate on the production of 
hadrons in this region.

We recall that, already in 1970s, 
it has become well-known 
(See, e.g., [\ref{Ochs78}, \ref{Kit81}] and the papers cited there.)
that inclusive longitudinal momentum distribution of a 
given type of hadron which has 
a valence quark in common with one of 
the colliding hadrons reflects directly 
the momentum distribution of 
this valence quark in that colliding hadron. 
Different experiments (See, e.g., [\ref{Gia79}-\ref{CHLM78}]) 
have shown that the longitudinal momentum distributions of the 
produced hadrons in the fragmentation region are very much 
similar to those of the corresponding valence quarks in the 
colliding hadrons. 
More precisely, it has been observed$^{\ref{Gia79}-\ref{CHLM78}}$ 
that, e.g., 
the number density of $\pi^+$ or that of $K^+$ in $pp$-collisions  
is proportional to the $u$-valence-quark distribution in proton; 
that of $\pi^-$ is proportional to that of $d$-valence quark. 
But that for $K^-$, which does not share a valence quark with the 
colliding hadron, drops much more fast and earlier than that 
for $K^+$ in the large $x_F$ region. 
These experimental facts suggest that, 
if there is a scattering process at all 
which takes place between these valence quarks
and other constituents of the colliding hadrons 
before they hadronize into hadrons,  
this scattering should {\it not} destroy 
the momentum distribution of the valence quarks. 
In other words, the momentum transfer in the 
scattering {\it cannot} be large. 
This implies that there should be no
hard scattering in these processes between  
such valence quarks and other constituents 
of the colliding hadrons.

Having these in mind, one is naturally led to the 
following picture$^{\ref{LM92}-\ref{Boros96}}$ 
for the production of mesons in 
the fragmentation regions: 
A valence quark of one of the colliding hadrons 
picks up an anti-sea-quark associated with the other 
and combine with it to form a meson 
which can be observed experimentally, 
e.g. $q_v^P+\bar q_s^T\to M$.
[Here, the subscript of a quark, $v$ or $s$, 
denotes whether it is for valence or sea quarks; 
the superscript, $P$ or $T$, denotes whether 
it is from the projectile or the target.]
This mechanism for hadron production 
is referred as ``direct formation'' or 
``direct fusion''. 

We note that the ``direct fusion'' mechanism mentioned 
above is very similar to 
the recombination model proposed by Das and Hwa\cite{Das77} 
several years ago. 
These are models which aim to describe the production 
of hadrons {\it in the fragmentation regions}.
This is also the simplest model 
which reproduces the data for hadron production in 
fragmentation regions of unpolarized 
hadron-hadron collisions.
There exist surely many other hadronization models 
in the literature, some of which can also reproduce  
most of the data for hadron production in the unpolarized 
hadronic reactions. 
Since our purpose here is to investigate the origin of the 
left-right asymmetries observed in the fragmentation region 
in single-spin hadron-hadron collisions, 
in particular the contribution of orbital motion 
of the valence quarks to these asymmetries, 
it is our intention to simplify the other factors as 
much as we are allowed in order to show the effects 
from the key factors. 
We therefore chose this model for the production 
of hadrons in the fragmentation region. 
In the following of this subsection, we will briefly 
review the picture and its main success in 
describing the unpolarized data to show 
that it can indeed describe the main properties of 
the hadrons in fragmentation regions.
For a review of the recombination models and/or 
comparison of this model to other hadronization models, 
see e.g. [\ref{HWA81},\ref{HWA87}].

For the sake of definiteness, we now consider 
the process $p(0)+p(0)\to M+X$. 
It is clear that, in this picture, 
the number density $N(x_F,M|s)$ of the produced mesons 
should be given by,
\begin{equation}
N(x_F,M|s)=N_0(x_F,M|s)+D(x_F,M|s),
\label{eq:NN0D}
\end{equation}
where $N_0(x_F,M|s)$ represents the contribution from non-direct
formation, which comes from the interaction of the seas 
(sea quarks, sea antiquarks and gluons) 
of the two colliding hadrons;
and $D(x_F,M|s)$ is the number density of the meson produced 
through the direct formation process $q_v^P+\bar q_s^T\to M$. 
Obviously, the $x_F$-dependence of $D$ can be obtained from 
the following integrals,
\begin{equation}
D(x_F,M \ | \ s) = \sum _{q_v,\bar q_s}
\int  dx^P dx^T q_v(x^P)
\bar q_s(x^T)K(x^P,q_v;x^T, \bar q_s| x_F, M,s). 
\label{eq:DfM}
\end{equation}
Here $q_v(x)$
is the distribution of the valence quarks
in the projectile proton, and $\bar q_s(x)$
is the sea-quark distribution in the target;
$K(x^P,q_v;x^T, \bar q_s| x_F, M,s)$ 
is the probability density for a valence quark 
of flavor $q_v$ with fractional momentum $x^P$ 
to combine directly with an anti-sea-quark of 
flavor $\bar q_s$ with fractional momentum $x^T$ 
to form a meson $M$ with fractional momentum $x_F$.
We note that whether, if yes how much, 
the $K$-function depends on
the dynamical details 
is something we do not know a priori.
But, what we do know is that 
this function has to guarantee 
the validity of all
the relevant conservation laws.
Hence, the simplest choice of the
corresponding $K$-function 
for e.g. $\pi^+=(u\bar d)$ is,
\begin{equation}
K(x^P,q_v;x^T, \bar q_s| x_F, \pi^+,s)=\kappa _\pi 
\delta _{q_v,u} \delta _{\bar q_s,\bar d} 
\delta (x^P-x_F) \delta (x^T-{x_0\over x_F}), 
\label{eq:Kpi+}
\end{equation}
where, $\kappa_\pi$ is a constant;  
the two Dirac-$\delta$-functions 
come from the energy and momentum
conservation which requires
$x^P\approx x_F$ and $x^T\approx x_0/x_F$ where $x_0=m^2/s$
($m$ is the mass of the produced meson).  
In this way, we obtain,
\begin{equation}
D(x_F,M|s)=\kappa q_v(x_F)\bar q_s({x_0\over x_F}),
\label{eq:DM}
\end{equation}

It can easily be seen that such a picture 
is consistent with the 
above mentioned experimental observations, namely 
in the fragmentation region, 
\begin{equation}
N(x_F,M|s)\sim D(x_F,M|s)\propto q_v(x_F).
\label{eq:Nlxf}
\end{equation}
It is also consistent with the existence 
\cite{Ben69,Bell73} of a limiting 
behavior in the fragmentation region at high energies. 
This is because, at extremely high energy, 
$x^T$ is very small and $\bar q_s(x^T) $ 
is very large. 
This implies that there exists  
a tremendously large number of antiseaquarks 
which are suitable to combine with the 
valence quarks $q_v^P$ of the projectile 
to form the mesons. 
Since we have only three valence quarks, 
further increasing of the energy, which means further 
increasing the number of the antiseaquarks, will 
bring nothing more,  
thus the distribution of the mesons produced in 
the fragmentation region remains the same.

\begin{figure}
\begin{center}
\psfig{file=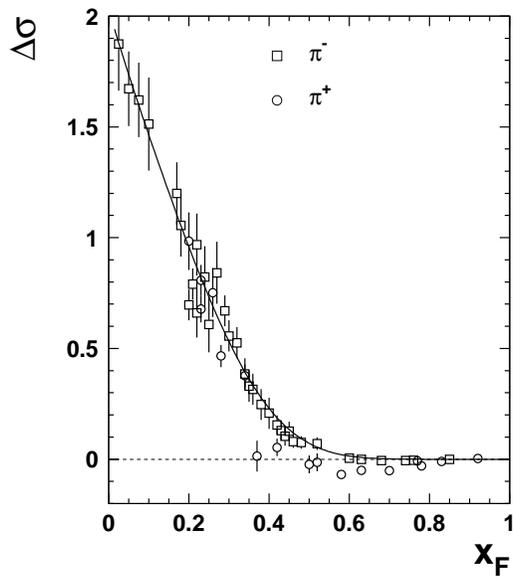,height=8cm}
\end{center}
\caption{The non-direct-formation parts of the inclusive 
pion production cross section for 
$p(0)+p(0)\to \pi^+ +X$ 
and $p(0)+p(0)\to \pi^- +X$ 
are shown as function of $x_F$.  
This figure is taken from [\protect\ref{LM94}]. }
\label{fig:dcspi}
\end{figure}

\begin{figure}
\begin{center}
\psfig{file=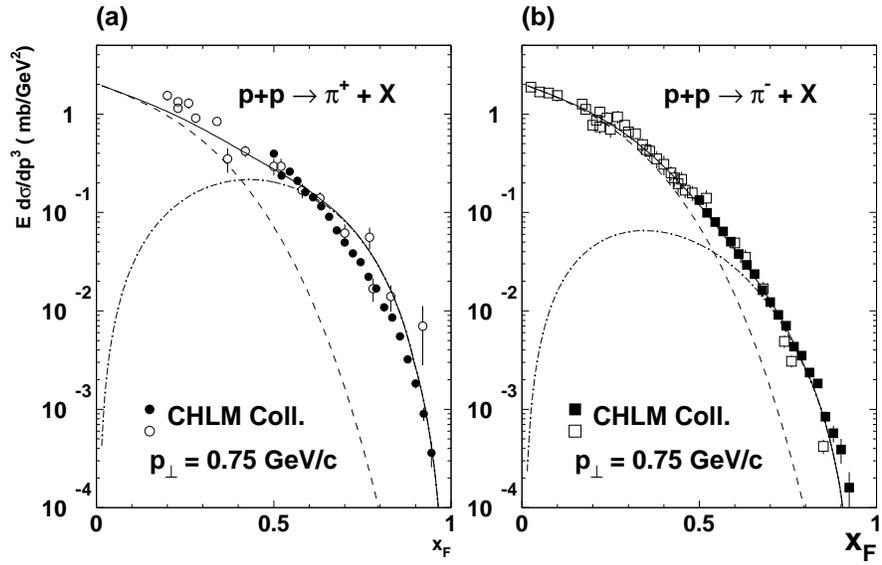,height=7cm}
\end{center}
\caption{The measured $x_F$-distribution for 
$p(0)+p(0)\to\pi^++X$ and that for $p(0)+p(0)\to\pi^-+X$
are shown as the sum of the following two parts: 
(1) the isospin-independent non-direct-formation part
(parameterization mentioned in Fig.2 shown as dashed curves), 
(2) the corresponding isospin-dependent parts
$\kappa _\pi u_v(x_F)\bar q_s(x_0/x_F) x_F$ for $\pi ^+$
and $\kappa _\pi d_v(x_F)\bar q_s(x_0/x_F) x_F$ for $\pi ^-$
(shown as dash-dotted curves).
This figure is taken from [\protect\ref{LM94}]. }
\label{fig:cspi}
\end{figure}

Since $N_0(x_F,M|s)$ comes from the interaction of the sea of the 
colliding hadrons, it is expected that 
$N_0$ should be isospin invariant 
and should be the same for particle and antiparticle.
This was first checked\cite{LM94} 
using data\cite{Gia79,CHLM73} 
for pion production. 
Here, one can obtain 
$N_0(x_F,M|s)$ by subtracting $D(x_F,M|s)$ from the 
corresponding data for 
$N(x_F,M|s)$, where $D(x_F,M|s)$ can be calculated 
by using the parameterizations 
for the quark distribution functions 
(See, e.g., [\ref{MT91},\ref{GRV92}]). 
The results for such a subtraction is shown in Fig.\ref{fig:dcspi}.
In Fig.\ref{fig:cspi},  we see 
the cross-sections\cite{Gia79,CHLM73} 
as the sums of two parts: 
the direct-formation part 
and the non-direct-formation part.  

Form these results, we not only see that 
the non-direct-formation part is indeed isospin-independent 
but also explicitly see the existence of 
a transition region in the inclusive cross-sections
near $x_F=0.4 \sim 0.5$.
In this region, $N_0(x_F,\pi|s)$ and 
$D(x_F,\pi|s)$ are expected to switch their roles:
While the former is the main contribution 
for $x_F<0.4\sim 0.5$, the latter 
begins to dominate for larger values of $x_F$.

\begin{figure}
\psfig{file=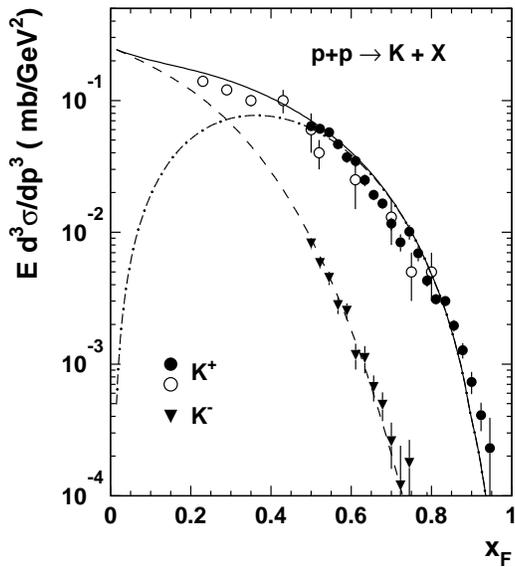,height=8cm}
\caption{Inclusive invariant cross section 
$Ed^3\sigma/d^3p$ for 
$p(0)+p(0)\to K^++X$ as a function of $x_F$ the ISR energies 
is shown as the sum of the following two parts: 
(1) the isospin-independent non-direct-formation part
which is taken as the same as 
$Ed^3\sigma/d^3p$ for $p(0)+p(0)\to K^-+X$ 
[parameterized as $N(1-x_F)^3exp(-10x_F^3)$, 
shown by the dashed curve], 
(2) the corresponding flavor-dependent 
direct formation part
$\kappa _K u_v(x_F)\bar s_s(x_0/x_F) x_F$ 
(shown by the dashed-dotted curve).
This figure is taken from [\protect\ref{BLM96}]. }
\label{fig:csK}
\end{figure}

There is an advantage to use $K$-production 
in $p(0)+p(0)\to K+X$ in testing this picture. 
Here, direct fusion of the valence quarks 
with suitable anti-sea-quarks does not 
contribute to the production of 
$K^-$ and $\bar K^0$.  
This means, for such kaons, we have only the 
contributions from the non-direct-formation parts $N_0(x_F,K|s)$, 
which should be the same for $K^\pm, K^0$ and $\bar K^0$.
Hence, $N_0(x_F,K|s)$ can be determined unambiguously by experiments. 
We thus obtain from the following relations 
between the number densities 
(or the corresponding differential cross sections) 
for $K$ produced in $p(0)+p(0)\to K+X$:
\begin{equation}
N(x_F,K^-|s)=N(x_F,\bar K^0|s)=N_0(x_F,K|s),
\label{eq:N0K}
\end{equation}
\begin{equation}
N(x_F,K^+|s)=N(x_F,K^-|s)+\kappa _K u_v(x_F)\bar s_s({x_0\over x_F}),
\label{eq:NK+}
\end{equation}
\begin{equation}
N(x_F,K^0|s)=N(x_F,K^-|s)+\kappa _K d_v(x_F)\bar s_s({x_0\over x_F}).
\label{eq:NK0}
\end{equation}
These are direct consequences of the picture 
and can be used to test the picture in a quantitative manner. 
A comparison of Eq.(\ref{eq:NK+}) 
with the ISR data\cite{CHLM73,CHLM78}
was made in [\ref{BLM96}] and is 
shown in Fig.\ref{fig:csK}.  
We see that the agreement is indeed very good.

The picture has also been applied\cite{BL96} to $\Lambda$ production. 
Here, we have the following three possibilities\cite{BL96}  
for direct formations of $\Lambda$
in $p(0)+p(0)\to \Lambda +X$:

(a) A $(u_vd_v)$-valence-diquark 
from the projectile $P$ picks up 
a $s_s$-sea-quark associated with the target 
$T$ and forms a $\Lambda $: 
$(u_vd_v)^P+s_s^T\to \Lambda $. 

(b) A $u_v$-valence-quark 
from the projectile $P$ picks up a 
$(d_ss_s)$-sea-diquark associated 
with the target $T$ and forms a $\Lambda $: 
$u_v^P+(d_ss_s)^T\to \Lambda $.

(c) A $d_v$-valence-quark 
from the projectile $P$ picks up a 
$(u_ss_s)$-sea-diquark associated with 
the target $T$ and forms a $\Lambda $: 
$d_v^P+(u_ss_s)^T\to \Lambda $.

Just as that for meson, 
the corresponding number densities 
are given by,
\begin{equation}
D_a(x_F,\Lambda |s)=\kappa_{\Lambda}^d  
f_D(x^P|u_vd_v) s_s(x^T),
\label{eq:DaLam}
\end{equation}
\begin{equation}
D_b(x_F,\Lambda |s)= \kappa_{\Lambda} 
   u_v(x^P) f_D(x^T|d_ss_s),
\label{eq:DbLam}
\end{equation}
\begin{equation}
D_c(x_F,\Lambda |s)= \kappa_{\Lambda} 
d_v(x^P) f_D(x^T|u_ss_s), 
\label{eq:DcLam}
\end{equation}
respectively. 
Here, $x^P\approx x_F$ and $x^T\approx m_\Lambda ^2/(sx_F)$, 
followed from energy-momentum conservation. 
$f_D(x|q_iq_j)$ is the diquark distribution functions, 
where $q_iq_j$ denotes the flavor and whether they are 
valence or sea quarks.
$\kappa _\Lambda ^d$ and $\kappa _\Lambda $ are two constants. 
The number density for $\Lambda$ is given by:
\begin{equation}
N(x_F,\Lambda |s)=N_0(x_F,\Lambda |s) +\sum_{i=a,b,c}D_i(x_F,\Lambda |s).
\label{eq:NDLam}
\end{equation}
where $N_0(x_F,\Lambda|s)$ is the non-direct-formation part. 

\begin{figure}
\begin{center}
\psfig{file=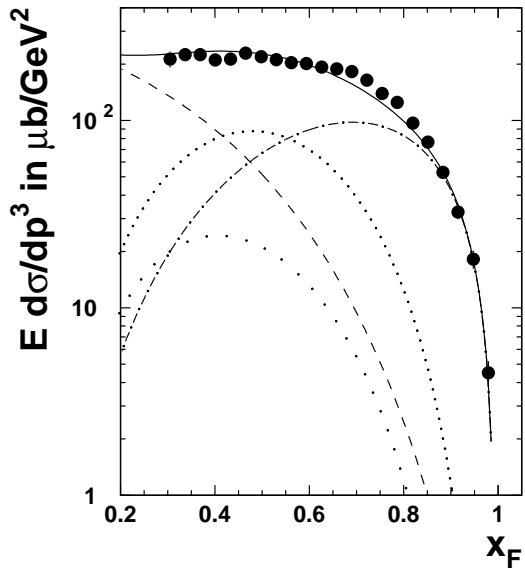,height=8cm}
\end{center}
\caption{The differential cross section $E d\sigma /dp^3$ for
$p(0)+p(0)\to \Lambda + X$
as a function of $x_F$
at $p_\perp = 0.65$ GeV/c and ISR-energies
as a sum of different contributions.
The dash-dotted and the two dotted lines 
represent the contributions from the 
direct formation processes (a), (b) and
(c) given in the text respectively.
The dashed line represents the
non-direct-formation part, which is parameterized as
$300(1-x_F)^2e^{-3x_F^3}$.
The solid line is the sum of all contributions.
The data is taken from [\protect\ref{Bob83}].
This figure is taken [\protect\ref{BL96}]. }
\label{fig:csLam}
\end{figure}

$D_i(x_F,\Lambda |s)$ has been calculated\cite{BL96} 
using the parameterizations of 
the quark and diquark distribution 
(See, e.g., [\ref{MT91},\ref{GRV92}] and [\ref{Dug93}]).
The two constants $\kappa _\Lambda $ and $\kappa _\Lambda ^d$ 
were fixed by fitting two data points 
in the large $x_F$-region. 
The results are compared with the data\cite{Bob83} in Fig.\ref{fig:csLam}.
We see that the data 
can indeed be fitted very well in the fragmentation region. 
In fact, compared with those for pions 
(See, e.g., [\ref{Gia79}] and [\ref{CHLM73}]), 
the characteristic feature of the data for $\Lambda$ 
(See, e.g., [\ref{Bob83}]) 
is that it is much broader than the latter 
in the large $x_F$ region, 
and this is just a direct consequence of the contribution 
from the valence diquarks through
the process (a) given above. 
We see also that the whole $0<x_F<1$ region 
can be divided into three parts: 
In the large $x_F$-region (say, $x_F> 0.6$), 
the direct process (a) plays the dominating role; 
and for small $x_F$-values ($x_F< 0.3$, say), 
the non-direct-formation part $N_0(x_F,\Lambda|s)$ 
dominates, while in the middle 
(that is, in the neighborhood of $x_F\sim 0.4-0.5$), 
the direct formation processes (b) and (c) 
provide the largest contributions.

\subsection{Left-right asymmetry for meson production}

Having seen that $A_N$ is 
significant only in the 
fragmentation region, 
that hadron production in this region can be 
well described in terms of the abovementioned 
direct fusion of the valence quarks with suitable 
antiseaquarks, and that one of the most remarkable properties 
of the valence quarks in polarized nucleon is that 
they have to perform orbital motions even when they are in their 
ground state,   
it is natural to ask: 
Can we describe the observed asymmetry data 
if we take the orbital motion of the valence quarks into account?
In particular, we have seen that 
the longitudinal momentum distribution  
of the produced hadrons in the fragmentation regions 
directly reflects the longitudinal momentum distributions 
of the valence quarks. 
Does the transverse momentum distribution 
of these hadrons reflect also the 
intrinsic transverse motion of the valence quarks 
in hadrons? 

It has been shown$^{\ref{Meng91}-\ref{BLM96}}$ 
that these questions should be 
answered in the affirmative. 
The existence of the left-right asymmetries in 
the fragmentation regions should be considered as a 
strong evidence for the existence of orbital motion of the 
valence quarks in a transversely polarized nucleon 
and the hadronic surface effects in 
single-spin hadron-hadron collisions. 
The arguments for the existence of the 
surface effect caused by the ``initial state interactions'' 
of the constituents of the two colliding hadrons when they are 
overlap in space are summarized in the following\cite{BLM93,BLM95}.

Hadrons --- the constituents of which are known to be
quarks and gluons --- are spatially extended objects.
Inside a hadron, the constituents interact with one another
through color forces,
and only color-singlets can leave the color-fields.
As spatially extended objects, the geometrical aspect of
hadrons plays an important
role in describing unpolarized high-energy hadron-hadron
collisions. In this connection, it is useful to recall the
following: Without specifying the details
about the constituents and their interactions, the
geometrical picture in which the colliding hadrons are simply
envisaged\cite{Ben69} 
to ``go through each other with attenuation
thereby in general get excited and break up"
is already able to describe
the characteristic limiting behaviors of
the final state particles in the fragmentation regions.
Having these in mind, we now examine
the role played by geometry in
direct-meson-formation processes
in the fragmentation regions when one of the
colliding hadrons is transversely polarized

For the sake of convenience and definiteness,
we consider the process
in the rest frame of the projectile $P$,
in which $P$ is an upwards transversely
polarized proton, and the target proton
$T$ is unpolarized.
We concentrate our attention on the transverse momentum distribution
of the directly formed mesons observed in the
projectile fragmentation region, and ask
the following questions:
(A) What do we expect to see if the observed
meson is formed through fusion of a
$u$-valence-quark, which was upwards polarized 
before the collision takes place, 
of $P$ near its front surface
(towards the target proton $T$) and
a {\it suitable} anti-sea-quark of $T$?
(B) What do we expect to see when this direct
fusion takes place elsewhere in $P$ ---
in particular near $P$'s back surface?

To answer these questions, we recall 
that, in the transversely polarized $P$,
the valence quarks are performing
orbital motion and the direction of such orbital motion
is determined by the polarization. 
In particular, an upwards polarized $u$-valence quark
is ``going-left" when looked near the front surface of $P$.
Hence, because of  momentum conservation,
the answer to (A) is the following:
Due to orbital motion of the valence quarks, 
it is more probable that 
this meson acquires an extra transverse momentum going to
the {\it left}. This is to be compared
with the direct fusion processes 
which take place when $T$ has already entered
$P$ after some time, in particular when
$T$ would be already near the back surface and about to leave $P$ 
if both $P$ and $T$ would remain undestroyed 
by the color interactions between their constituents. 
We note: before $T$ enters $P$,
axial symmetry w.r.t.
the polarization axis requires that
the upwards $u$-valence quark is 
``right-going" near the back surface of $P$. 
Hence, if such a valence $u$-quark could retain
its transverse momentum until
$T$ reaches $P$'s back surface, 
the produced meson would have the same probability 
to go right as that for the meson 
formed near the front surface to go left. 
We would obtain that, in total, the chance for a meson 
formed by fusion of an upwards polarized $u$ valence quark 
with suitable anti-sea-quark to go left and 
that to go right equal to each other. 
But, since color forces between the constituents
in the $P-T$ system become effective as soon as
$P$ and $T$ begin to overlap,
and such interactions 
(often referred as ``initial state interaction'') 
in general cause changes in the
intrinsic motion of the constituents inside the hadrons $P$ and $T$.
The above-mentioned $u$-valence-quark in $P$
would be able to retain
its initial momentum {\it only  if } 
the time needed for $T$ to travel through $P$
would be {\it less} than the time needed for the color interaction
to propagate the same distance.
Hence, in a picture --- in fact in any picture ---
in which $T$ {\it cannot} go through $P$
{\it without} any resistance/attenuation, while color gluons propagate with
the velocity of light,
it is more likely that {\it the contrary is true}. 
This means, the answer to Question (B) should be:
If the direct fusion takes place near the back surface of $P$, 
it is very likely that the $u$ valence quark has already 
lost the information of polarization which it had 
in the initial state hadron thus the produced hadron 
can go left or right with equal probabilities.
We thus obtain that meson formed through direct fusion 
of an upwards polarized valence quark with a suitable 
anti-sea-quark has a large probability to go left.

Together with the surface effect, the orbital motion 
of a polarized valence quark in a polarized nucleon 
leads to a correlation between 
the polarization of the valence quarks 
and the direction of transverse motion 
of the mesons produced through the direct fusion of 
the valence quarks and suitable antiquarks. 
More precisely, we obtain that
mesons produced through the direct formation 
of upwards transversely polarized valence quarks 
of the projectile 
with suitable antiseaquarks associated with the target 
have large probability to go left and vice versa. 
Hence, once we know the polarization of the projectile, 
we can use the baryon wave function 
to determine the polarization 
of its valence quarks and 
the signs of $A_N$'s for the produced hadrons. 
E.g., for $p(\uparrow )+p(0)\to \pi +X$, 
we obtain the results in table \ref{tab:ANpptopi}.
Here, in the table, the following coordinate is used: 
The projectile is moving in $+z$ direction, 
polarization up is $+x$ direction; 
therefore moving to the left means $p_y<0$ 
(denoted by $\leftarrow$) 
and moving to the right means $p_y>0$ 
(denoted by $\rightarrow$).  
From this table we see clearly the following:
\begin{itemize}
\itemsep=-0.10truecm
\item \vskip -0.2truecm $A_N[\ p(\uparrow )+p(0)\to \pi ^++X] >0,$ 
\ \ \ $A_N[\ p(\uparrow )+p(0)\to \pi ^-+X] <0,$ 
\item $0<A_N[\ p(\uparrow )+p(0)\to \pi ^0+X] <A_N[\ p(\uparrow )+p(0)\to \pi ^++X] $,
\item The magnitudes of all these $A_N$'s increase with increasing
$x_F$. 
\end{itemize}
\vskip -0.12truecm \noindent
{\it All these qualitative features are 
in good agreement with the data}$^{\ref{Kle76}-\ref{E70496b}}$.

\begin{table}
\caption{Properties of $\pi^{\pm },\ \pi^0$ or $\eta $ 
in $p(\uparrow )+p(0)\to \pi ^{\pm }({\rm or}\ \pi^0, \eta )+X$}
\label{tab:ANpptopi}
\centering
{\small
\begin{tabular}{l@{\extracolsep{0.6truecm}}
  c@{\extracolsep{0.6truecm}}c@{\extracolsep{0.6truecm}}
  c@{\extracolsep{0.6truecm}}c@{\extracolsep{1.0truecm}}
 l@{\extracolsep{0.6truecm}}
  c@{\extracolsep{0.6truecm}}c@{\extracolsep{0.6truecm}}
  c@{\extracolsep{0.6truecm}}
  c@{\extracolsep{0.6truecm}}c@{\extracolsep{0.6truecm}}
  c@{\extracolsep{0.6truecm}}}
\hline \\[-0.3truecm]
\multicolumn{5}{l} {$P$(sea)---$T$(val)} & 
\multicolumn{7}{l} {$P$(val)---$T$(sea)} \\[0.2truecm] 
\hline \\[-0.3truecm]
$P$(sea)&$u$ &$\bar u$ &$d$ &$\bar d$ &
 $P$(val)&\multicolumn{2}{c}{$u$}      &  &\multicolumn{2}{c}{$d$}    & \\ 
$p_y$   &0   &0        &0   &0        &
 $p_y$   &$\leftarrow$ &$\rightarrow $&  &$\leftarrow$&$\rightarrow$& \\
Weight  &1   &1        &1   &1        &
 Weight  &5/3          &1/3           &  &1/3         &2/3          &
                                                                  \\[0.2truecm]
\hline\\[-0.3truecm]
$T$(val)& &$d$       &  &$u$      &  
  $T$(sea)&\multicolumn{2}{c}{$\bar d$} &$d$&\multicolumn{2}{c}{$\bar u$}&$u$\\ 
$p_y$   & &0         &  &0        &  
  $p_y$   &\multicolumn{2}{c}{0}        &0  &\multicolumn{2}{c}{0}       &0 \\ 
Weight  & &1         &  &2        &  
  Weight  &\multicolumn{2}{c}{1}        &1  &\multicolumn{2}{c}{1}       &1 
                                                               \\[0.1truecm] 
Product & &$d\bar u$ &  &$u\bar d$& 
  Product &\multicolumn{2}{c}{$u\bar d$}&   &\multicolumn{2}{c}{$d\bar u$}& \\ 
$p_y$   & &0         &  &0        & 
  $p_y$   &$\leftarrow$ &$\rightarrow$ &   &$\leftarrow$ &$\rightarrow$ & \\
Weight  & &1         &  &2        &
  Weight  &5/3          &1/3           &   &1/3          &2/3           &  
                                                                   \\[0.2truecm]
\hline\\[-0.3truecm]
$T$(val)& &$u$       &  &$d$      &  
  $T$(sea)&\multicolumn{2}{c}{$\bar u$} &$u$&\multicolumn{2}{c}{$\bar d$}&$d$\\ 
$p_y$   & &0         &  &0        &  
  $p_y$   &\multicolumn{2}{c}{0}        &0  &\multicolumn{2}{c}{0}       &0 \\ 
Weight  & &2         &  &1        &  
  Weight  &\multicolumn{2}{c}{1}        &1  &\multicolumn{2}{c}{1}       &1 
                                                                 \\[0.1truecm] 
Product & &$u\bar u$ &  &$d\bar d$& 
  Product &\multicolumn{2}{c}{$u\bar u$}&   &\multicolumn{2}{c}{$d\bar d$}& \\ 
$p_y$   & &0         &  &0        & 
  $p_y$   &$\leftarrow$ &$\rightarrow$ &   &$\leftarrow$ &$\rightarrow$ & \\
Weight  & &2         &  &1        &
  Weight  &5/3          &1/3           &   &1/3          &2/3           &  
                                                         \\[0.2truecm]\hline 
\end{tabular}
   }
\end{table}

Similar tables can also be constructed in a straightforward manner 
for the production of other mesons and/or in processes using other 
projectiles and/or targets.
In this way, we obtained many other direct associations 
of the picture which can be tested directly by experiments.
E.g.: 

(A) If we use, instead of $p(\uparrow)$, 
$\bar p(\uparrow)$-projectile, we obtain table \ref{tab:ANpbarptopi}.
Compare the results in this table with those in table \ref{tab:ANpptopi},
we see that, while $A_N$ for $\pi^0$ remains the same, 
those for $\pi^+$ and $\pi^-$ change their roles.
More precisely, $A_N$ for $\bar p(\uparrow )+p(0)\to \pi^-+X$ should be  
approximately the same as that for
$p(\uparrow )+p(0)\to \pi^++X$; 
that for $\bar p(\uparrow )+p(0)\to \pi^++X$ should be  
approximately the same as that for
$p(\uparrow )+p(0)\to \pi^-+X$. 

(B) It is also clear that the asymmetries 
are expect to be significant in the fragmentation of the polarized  
nucleon but should be absent in the fragmentation of the unpolarized hadron.
This implies in particular that 
there should be no asymmetry 
in the beam fragmentation region of
$\pi^-+p(\uparrow )\to \pi+X$. 

(A) is a prediction of [\ref{Meng91}, \ref{LM92}] and [\ref{BLM93}] 
and has been confirmed by the E704 data\cite{Yok92,E70496b}.
(B) was also predicted in [\ref{Meng91}] and [\ref{BLM93}] 
and has been confirmed by the data\cite{Apo90}.

\begin{table}
\caption{Properties of $\pi^{\pm },\ \pi^0$ or $\eta $ 
in $\bar p(\uparrow )+p(0)\to \pi ^{\pm }({\rm or}\ \pi^0, \eta )+X$}
\label{tab:ANpbarptopi}
\centering
{\small
\begin{tabular}{l@{\extracolsep{0.6truecm}}
  c@{\extracolsep{0.6truecm}}c@{\extracolsep{0.6truecm}}
  c@{\extracolsep{0.6truecm}}c@{\extracolsep{1.0truecm}}
 l@{\extracolsep{0.6truecm}}
  c@{\extracolsep{0.6truecm}}c@{\extracolsep{0.6truecm}}
  c@{\extracolsep{0.6truecm}}
  c@{\extracolsep{0.6truecm}}c@{\extracolsep{0.6truecm}}
  c@{\extracolsep{0.6truecm}}}
\hline \\[-0.3truecm]
\multicolumn{5}{l} {$P$(sea)---$T$(val)} & 
\multicolumn{7}{l} {$P$(val)---$T$(sea)} \\[0.2truecm] 
\hline \\[-0.3truecm]
$P$(sea)&$u$ &$\bar u$ &$d$ &$\bar d$ &
 $P$(val)&\multicolumn{2}{c}{$\bar u$}   &  &\multicolumn{2}{c}{$\bar d$}& \\ 
$p_y$   &0   &0        &0   &0        &
 $p_y$   &$\leftarrow$ &$\rightarrow $&  &$\leftarrow$&$\rightarrow$     & \\
Weight  &1   &1        &1   &1        &
 Weight  &5/3          &1/3           &  &1/3         &2/3          &
                                                                  \\[0.2truecm]
\hline\\[-0.3truecm]
$T$(val)& &$d$       &  &$u$      &  
  $T$(sea)&\multicolumn{2}{c}{$d$}&$\bar d$&\multicolumn{2}{c}{$u$}&$\bar u$\\ 
$p_y$   & &0         &  &0        &  
  $p_y$   &\multicolumn{2}{c}{0}  &0       &\multicolumn{2}{c}{0}  &0       \\ 
Weight  & &1         &  &2        &  
  Weight  &\multicolumn{2}{c}{1}  &1       &\multicolumn{2}{c}{1}  &1 
                                                                 \\[0.1truecm] 
Product & &$d\bar u$ &  &$u\bar d$& 
  Product &\multicolumn{2}{c}{$d\bar u$} & &\multicolumn{2}{c}{u$\bar d$}&  \\ 
$p_y$   & &0         &  &0        & 
  $p_y$   &$\leftarrow$ &$\rightarrow$ &   &$\leftarrow$ &$\rightarrow$ & \\
Weight  & &1         &  &2        &
  Weight  &5/3          &1/3           &   &1/3          &2/3           &  
                                                                   \\[0.2truecm]
\hline\\[-0.3truecm]
$T$(val)& &$u$       &  &$d$      &  
  $T$(sea)&\multicolumn{2}{c}{$u$}&$\bar u$&\multicolumn{2}{c}{$d$}&$\bar d$\\ 
$p_y$   & &0         &  &0        &  
  $p_y$   &\multicolumn{2}{c}{0}        &0  &\multicolumn{2}{c}{0}       &0 \\ 
Weight  & &2         &  &1        &  
  Weight  &\multicolumn{2}{c}{1}        &1  &\multicolumn{2}{c}{1}       &1 
                                                                 \\[0.1truecm] 
Product & &$u\bar u$ &  &$d\bar d$& 
  Product &\multicolumn{2}{c}{$u\bar u$}&   &\multicolumn{2}{c}{$d\bar d$}& \\ 
$p_y$   & &0         &  &0        & 
  $p_y$   &$\leftarrow$ &$\rightarrow$ &   &$\leftarrow$ &$\rightarrow$ & \\
Weight  & &2         &  &1        &
  Weight  &5/3          &1/3           &   &1/3          &2/3           &  
                                                         \\[0.2truecm]\hline 
\end{tabular}
   }
\end{table}

Since, as have seen in the last section, the 
$x_F$-dependence for the number density of mesons 
for direct formation can be calculated easily,  
we can also calculate the $x_F$-dependence for $A_N$ 
in a quantitative manner.
Such calculations have been carried out in 
[\ref{LM92}--\ref{BL96}]. 
We summarize them in the following.
For the sake of explicity, we 
consider $p(\uparrow )+p(0)\to M+X$.
We recall that [see Eqs.(1-3)] 
$A_N(x_F,M|s)$ is defined as 
the ratio of the difference and the sum 
of $N_L(x_F,M|s,\uparrow )$ and 
$N_L(x_F,M|s,\downarrow )$. 
We denote by $D(x_F,M,+|s,\uparrow )$ 
the number density of $M$'s produced through direct fusion
of the valence quarks which are polarized in the same direction 
as the transversely polarized proton and 
$D(x_F,M,-|s,\uparrow )$ the 
corresponding number density for $M$'s 
formed by the valence quarks polarized in 
the opposite direction as the projectile proton.
Then the $N$'s can be written as:
\begin{equation}
N_L(x_F,M|s,\uparrow )=\alpha D(x_F,M,+|s,\uparrow )+
(1-\alpha) D(x_F,M,-|s,\uparrow )+N_{0L}(x_F,M|s), 
\label{eq:NlM}
\end{equation}
\begin{equation}
N_L(x_F,M|s,\downarrow )=(1-\alpha) D(x_F,M,+|s,\downarrow )+
\alpha  D(x_F,M,-|s,\downarrow )+N_{0L}(x_F,M|s). 
\label{eq:NrM}
\end{equation}
Here, $\alpha$ stands for the probability for a meson 
produced by the direct fusion of an upwards polarized 
valence quark with a suitable anti-sea-quark to go left.
It follows from Eqs.(\ref{eq:NlM}) and (\ref{eq:NrM}), 
\begin{equation}
N_L(x_F,M |s,\uparrow)-N_L(x_F,M|s,\downarrow)= C\ 
          [ D(x_F,M,+|s,tr)-D(x_F,M, -|s,tr)],
\label{eq:dNlrM}
\end{equation}
where $C\equiv 2\alpha -1$, 
$D(x_F,M,\pm|s,tr)\equiv 
 D(x_F,M,\pm|s,\uparrow )=D(x_F,M,\pm|s,\downarrow\nobreak)$. 
Hence, 
\begin{equation}
A_N(x_F,M|s)={C\Delta D(x_F,M|s,tr) \over 2N_{0L}(x_F,M|s)+D(x_F,M|s)},
\label{eq:ANm}
\end{equation}
Since mesons directly 
formed through fusion
of upwards polarized valence-quark of the projectile and 
anti-sea-quarks of the target 
have a larger probability to go left, 
we have $1/2<\alpha <1$ which implies $0<C<1$.
It can easily be seen that 
$N_0(x_F,M|s)[= 2N_{0L}(x_F,M|s)]$ 
--- especially the interplay 
between this quantity and the corresponding $D(x_F,M|s)$
--- plays a key role 
in understanding the $x_F$-dependence of $A_N(x_F,M|s)$.   

As we have shown in last section, $N_0(x_F,M|s)$ 
can be obtained from the unpolarized data.
The $D$'s can be obtained from the following integrals,
\begin{equation}
D(x_F,M,\pm \ | \ s,tr) = \sum _{q_v,\bar q_s}
\int  dx^P dx^T q_v^{\pm}(x^P|tr)
\bar q_s(x^T)K(x^P,q_v;x^T, \bar q_s| x_F, M,s). 
\label{eq:DMpm}
\end{equation}
Here $q^\pm _v(x|tr)$
is the distribution of the valence quarks
polarized in the same or in the opposite direction of the
transversely polarized proton.  
Using the $K(x^P,q_v;x^T, \bar q_s| x_F, M,s)$'s 
as that for $\pi^+$ given in Eq.(\ref{eq:Kpi+}), 
we obtain the $D(x_F,M,\pm|s,tr)$'s and thus, 
\begin{equation}
A_N(x_F,\pi^+|s)={C\kappa_\pi\Delta u_v(x_F|tr) \bar d_s({x_0\over x_F}) 
\over N_0(x_F,\pi|s)+\kappa_\pi u_v(x_F) \bar d_s({x_0\over x_F}) },
\label{eq:Anpi+}
\end{equation}
\begin{equation}
A_N(x_F,\pi^-|s)={C\kappa_\pi\Delta d_v(x_F|tr) \bar u_s({x_0\over x_F}) 
\over N_0(x_F,\pi|s)+\kappa_\pi d_v(x_F) \bar u_s({x_0\over x_F}) },
\label{eq:Anpi-}
\end{equation}
\begin{equation}
A_N(x_F,\pi^0|s)={C\kappa_\pi
 [\Delta u_v(x_F|tr) \bar u_s({x_0\over x_F})+
\Delta d_v(x_F|tr) \bar d_s({x_0\over x_F})]
\over 2N_0(x_F,\pi|s)
+\kappa_\pi [u_v(x_F) \bar u_s({x_0\over x_F})+
d_v(x_F) \bar d_s({x_0\over x_F})] },
\label{eq:Anpi0}
\end{equation}
\begin{equation}
A_N(x_F,K^+|s)={C\kappa_K\Delta u_v(x_F|tr) \bar s_s({x_0\over x_F}) 
\over N(x_F,K^-|s)+\kappa_K u_v(x_F) \bar s_s({x_0\over x_F}) },
\label{eq:AnK+}
\end{equation}
\begin{equation}
A_N(x_F,K^0|s)={C\kappa_K\Delta d_v(x_F|tr) \bar s_s({x_0\over x_F}) 
\over N(x_F,K^-|s)+\kappa_K d_v(x_F) \bar s_s({x_0\over x_F}) },
\label{eq:AnK0}
\end{equation}
and $A_N(x_F,K^-|s)=A_N(x_F,\bar K^0|s)=0$. 

It should be emphasized that the $q^{\pm}_v(x|tr)$ defined above is 
quite different from the $q^{\pm}(x|l)$ defined in the parton model: 
First, while the former refers to the transverse polarization, 
the latter refers to the longitudinal polarization. 
Second, the $\pm$ in the former refers to the 
third component of the total angular momenta of the valence quarks, 
that in the latter refers to the helicities of the quarks.
$q^{\pm}_v(x|tr)$ has not yet been determined experimentally. 
The only thing we know theoretically is that they have to 
satisfy the following constraints:
\begin{equation}
\int dx u^+_v(x|tr)=5/3,\ \ \
\int dx u^-_v(x|tr)=1/3,
\label{eq:u+-int}
\end{equation}
\begin{equation}
\int dx d^+_v(x|tr)=1/3,\ \ \ 
\int dx d^-_v(x|tr)=2/3,
\label{eq:d+-int}
\end{equation}
Hence, the simplest ansatz for these $q^\pm_v(x|tr)$ 
is the following:
\begin{equation}
u^+_v(x|tr)=(5/6)u_v(x),\ \ \ 
u^-_v(x|tr)=(1/6)u_v(x),
\label{eq:u+-u}
\end{equation}
\begin{equation}
d^+_v(x|tr)=(1/3)d_v(x),\ \ \ 
d^-_v(x|tr)=(2/3)d_v(x),
\label{eq:d+-d}
\end{equation}
This was used in [\ref{BLM93}-\ref{BLM96}] 
to make rough estimation of $A_N$.

\subsubsection{$A_N$ for $\pi$ production}

\begin{figure}
\begin{center}
\psfig{file=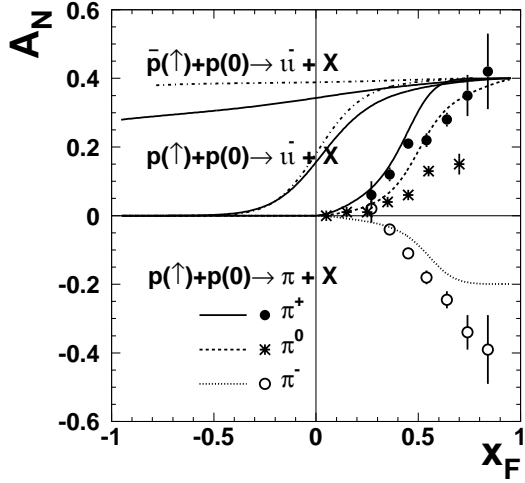,height=7cm}
\end{center}
\caption{Calculated results for 
$A_N$ as a function of $x_F$ 
for $p(\uparrow )+p(0)\to \pi +X$ and that for 
for $p(\uparrow )+p(0)\to ll +X$ or 
$\bar p(\uparrow )+p(0)\to ll +X$. 
The data are the E704 results for $\pi$ production 
which are the same as those shown in Fig.1. 
For lepton-pair, the solid 
and the dash-dotted lines are for 
$Q=4$ and 9GeV respectively. 
This figure is taken from [\protect\ref{BLM95}].} 
\label{fig:ANpipp}
\end{figure}

A rough estimation for $A_N$ for 
$p(\uparrow )+p(0)\to \pi+X$ was made 
in [\ref{BLM93}] using 
Eqs.(\ref{eq:Anpi+})--(\ref{eq:Anpi0}) and 
Eqs.(\ref{eq:u+-u})--(\ref{eq:d+-d}). 
The results are shown in Fig.\ref{fig:ANpipp}. 
We see that all the qualitative features of the data 
are well reproduced. 

It has also been found\cite{BLM95} that 
there exist many simple relations 
between the $A_N$'s for hadron production 
in reactions using different projectile-target 
combinations. 
E.g., 
\begin{equation}
A_N^{\bar p(\uparrow)p(0)}(x_F,\pi^+|s) \approx 
A_N^{p(\uparrow)p(0)}(x_F,\pi^-|s), 
\label{eq:ppbarpi+}
\end{equation}
\begin{equation}
A_N^{\bar p(\uparrow)p(0)}(x_F,\pi^-|s) \approx 
A_N^{p(\uparrow)p(0)}(x_F,\pi^+|s), 
\label{eq:ppbarpi-}
\end{equation}
\begin{equation}
A_N^{\bar p(\uparrow)p(0)}(x_F,\pi^0|s) \approx 
A_N^{p(\uparrow)p(0)}(x_F,\pi^0|s), 
\label{eq:ppbarpi0}
\end{equation}
\begin{equation}
A_N^{p(0)p(\uparrow)}(x_F,\pi^+|s) =
A_N^{p(0)n(\uparrow)}(x_F,\pi^-|s) \approx 
A_N^{\pi p(\uparrow)}(x_F,\pi^+|s), 
\end{equation}
\begin{equation}
A_N^{p(0)p(\uparrow)}(x_F,\pi^-|s) =
A_N^{p(0)n(\uparrow)}(x_F,\pi^+|s) \approx 
A_N^{\pi p(\uparrow)}(x_F,\pi^-|s), 
\end{equation}
\begin{equation}
A_N^{p(0)p(\uparrow)}(x_F,\pi^0|s) =
A_N^{p(0)n(\uparrow)}(x_F,\pi^0|s) \approx 
A_N^{\pi p(\uparrow)}(x_F,\pi^0|s), 
\end{equation}
\begin{equation}
A_N^{p(0)D(\uparrow)}(x_F,\pi^+|s) = 
A_N^{p(0)D(\uparrow)}(x_F,\pi^-|s) = 
A_N^{p(0)D(\uparrow)}(x_F,\pi^0|s) = 
A_N^{p(0)p(\uparrow)}(x_F,\pi^0|s). 
\end{equation}
Here, the superscript of $A_N$ denotes 
the projectile and the target of the reaction.
Eqs.(\ref{eq:ppbarpi+})-(\ref{eq:ppbarpi0}) are consistent 
with the data$^{\ref{E70488}-\ref{E70496b}}$.
In fact, Eqs.(\ref{eq:ppbarpi+}) and (\ref{eq:ppbarpi-}) 
were predicted\cite{Meng91,BLM93} before 
the corresponding were available.
The others can be tested by future experiments.

\subsubsection{$K$ production}

\begin{figure}
\begin{center}
\psfig{file=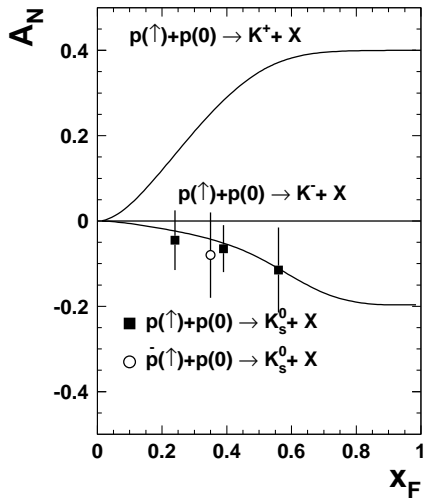,height=7cm}
\end{center}
\caption{Calculated results for 
$A_N$ as a function for $p(\uparrow)+p(0)\to K+X$. 
The data are the E704 preliminary results$^{15}$.
This figure is taken from [\protect\ref{BLM96}].}
\label{fig:ANKpp}
\end{figure}

Comparison of Eq.(\ref{eq:AnK+}) with Eq.(\ref{eq:Anpi+}), 
and Eq.(\ref{eq:AnK0}) with Eq.(\ref{eq:Anpi-}) explicitly 
shows\cite{BLM95,BLM96} 
that $A_N(x_F,K^+|s)$ should be similar to $A_N(x_F,\pi^+|s)$, and 
that $A_N(x_F,K^0|s)$ should be similar to $A_N(x_F,\pi^-|s)$. 
Since both the $K_S^0$ and $K_L^0$ are linear combinations 
of $K^0$ and $\bar K^0$, the left-right asymmetry should be the same 
for both of them, and it is given by\cite{BLM96}
\begin{equation}
A_N(x_F,K^0_S|s)={C\kappa_K\Delta d_v(x_F|tr) \bar s_s({x_0\over x_F}) 
\over 2N(x_F,K^-|s)+\kappa_K d_v(x_F) \bar s_s({x_0\over x_F})}.
\label{eq:AnK0s}
\end{equation}
Hence $A_N(x_F,K^0_S|s)$ should have the same sign 
as $A_N(x_F,\pi^-|s)$. This has been confirmed by the 
preliminary E704 data\cite{Bra95}.
A quantitative estimation of 
the $A_N$'s for Kaons have also been made\cite{BLM96}.
The results are shown in Fig.\ref{fig:ANKpp}.

In this connection, 
it may be particularly interesting to note the following. 
If we use, instead of transversely polarized proton beam,  
transversely polarized anti-proton beam, 
we have, 
\begin{equation}
A_N^{\bar p(\uparrow)p}(x_F,K^-|s)=
{C\kappa_K\Delta \bar u_v^{\bar p}(x_F|tr) s_s({x_0\over x_F}) 
\over N(x_F,K^-|s)+\kappa_K \bar u_v^{\bar p}(x_F) s_s({x_0\over x_F}) },
\label{eq:ANK-pbar}
\end{equation}
\begin{equation}
A_N^{\bar p(\uparrow)p}(x_F,K^0_S|s)=
{C\kappa_K\Delta \bar d_v^{\bar p}(x_F|tr) s_s({x_0\over x_F})
\over 2N(x_F,K^-|s)+\kappa_K \bar d_v^{\bar p}(x_F) s_s({x_0\over x_F}) },
\label{eq:ANK0spbar}
\end{equation}
and $A_N^{\bar p(\uparrow)p}(x_F,K^+|s)=0$. 
Using $\Delta \bar q_v^{\bar p}(x|tr)=\Delta q_v(x|tr)$
and $\bar q_v^{\bar p}(x)=q_v(x)$, 
we obtain that 
\begin{equation}
A_N^{\bar p(\uparrow)p}(x_F,K^-|s)=A_N(x_F,K^+|s),
\label{eq:ANK-ppbar}
\end{equation}
\begin{equation}
A_N^{\bar p(\uparrow)p}(x_F,K^0_S|s)=A_N(x_F,K^0_S|s).
\label{eq:ANK0ppbar}
\end{equation}
The latter has also been confirmed 
by the preliminary E704 data\cite{Bra95}.

\subsection{$A_N$ for lepton-pair production}

Not only mesons but also 
lepton pairs should\cite{BLM93} 
exhibit left-right asymmetry 
in single-spin processes. 
This is particularly interesting because 
the production mechanism 
--- the well known Drell-Yan mechanism\cite{DY70} --- 
is very clear, 
and there is no fragmentation hence 
no contribution from fragmentation.  
The measurements of such asymmetry 
provide a crucial test 
of the model and is also very helpful to 
study the origin of the observed $A_N$ in general. 
This will be discussed in more detail in Section 5.

\begin{figure}
\begin{center}
\psfig{file=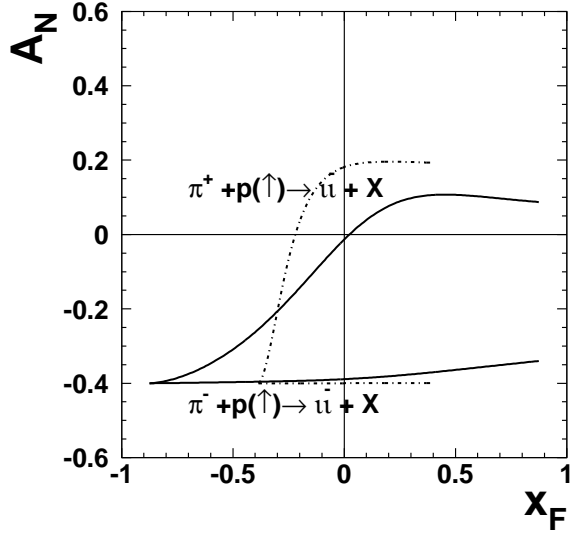,height=5.6cm}
\end{center}
\caption{$A_N$ as a function of $x_F$ for
$\pi+p(\uparrow) \to l\bar l+X$ calculated in [\ref{BLM95}]. 
at $p_{inc}=70$GeV/c. The solid and dashed-dotted lines are for 
$Q=4$ and 9GeV respectively. } 
\label{fig:ANllpip}
\end{figure}

\begin{figure}
\begin{center}
\psfig{file=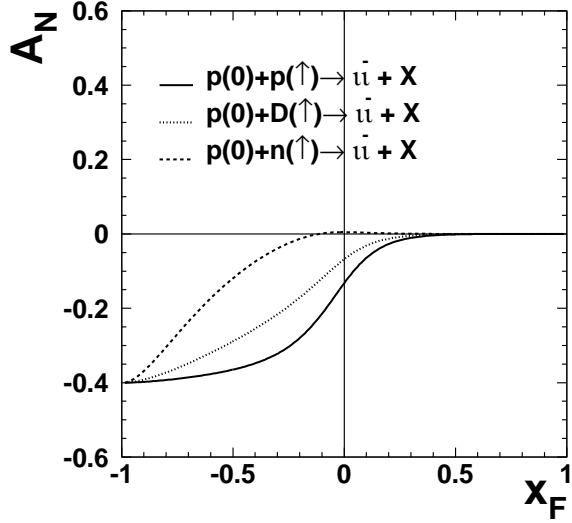,height=5.6cm}
\end{center}
\caption{$A_N$ as a function of $x_F$ for
 $p(0)+p(\uparrow)$ [or $n(\uparrow)$, or $D(\uparrow)$] 
 $\to l\bar l+X$ calculated in [\ref{BLM95}]. 
at $p_{inc}=820$GeV/c. The solid and dashed-dotted lines are for 
$Q=4$ and 9GeV respectively. } 
\label{fig:ANllppnD}
\end{figure}

The calculation for $A_N$ for lepton-pair 
is straightforward.  
Here, we can calculate not only the numerator 
in a similar way as that for 
hadron production, but also the denominator 
by using the Drell-Yan mechanism.
Such calculations have been done 
in [\ref{BLM93}] and [\ref{BLM95}]. 
Here, we give some examples of the results in 
Figs. \ref{fig:ANpipp}, \ref{fig:ANllpip} and \ref{fig:ANllppnD}.

\subsection{$A_N$ for hyperon production}

Striking $A_N$ 
have also been observed\cite{E70495} for $\Lambda $.
Compared with those for pions\cite{E70491}, 
$A_N$ for $\Lambda $ as a function of $x_F$ 
shows the following (C.f. Fig.1): 

(1) Similar to those for pions, 
$A_N(x_F,\Lambda|s )$ is significant 
in the large $x_F$ region ($x_F> 0.6$, say), 
but small in the central region. 

(2) $A_N(x_F,\Lambda |s )$ is negative 
in the large $x_F$ region, and 
it behaves similarly as 
$-A_N(x_F,\pi ^+|s)$ does in this region. 

(3) $A_N(x_F,\Lambda |s)$ begins to rise up 
later than $A_N(x_F,\pi^\pm |s)$ does. 
For $x_F$ in the neighborhood 
of $x_F\sim 0.4-0.5$, 
it is even positive.
In the region $0<x_F<0.5$, 
its behavior is similar to 
that of $A_N(x_F,\pi^0 |s)$.

At the first sight, these results, in particular 
those at large $x_F$, are rather surprising. 
This is because, as mentioned in section 4.2, 
the large $x_F$ region is dominated by the direct fusion 
process (a) $(u_vd_v)^P+s_s^T\to \Lambda$, 
where, according to the wave function of $\Lambda$, 
the $(u_vd_v)^P$ diquark from the projectile 
has to be in a spin-zero state. 
How can a spin zero object transfer the information of 
polarization to the produced $\Lambda$? 
This question has been discussed in [\ref{BL96}] 
and a solution has been suggested where 
associated production plays an important role. 
It has been pointed out\cite{BL96} that
the direct fusion process (a) should be 
predominately associated with the production of 
a meson directly formed through fusion of 
the remaining $(u_v^a)^P$ valence quark of the projectile 
with a suitable anti-sea-quark of the target. 
This $(u_v^a)^P$ carries the information of polarization 
of the projectile thus the associatively produced 
meson ($K^+$ or $\pi^+$) which contains this $(u_v^a)^P$ 
should have a large probability to 
obtain an extra transverse momentum to the left 
if the projectile is upwards polarized.  
This is caused by the 
the intrinsic transverse motion 
of the $u$-valence quark  and the surface effect.  
According to momentum conservation, 
the intrinsic transverse momentum of the valence quark 
should be approximately compensated 
by that of the other valence quarks.
Hence, the left valence-diquark thus  
the $\Lambda$ produced through 
the above mentioned process (a) should 
have a large probability to 
obtain an extra transverse momentum in the opposite 
direction as the associatively produced meson, 
i.e. to the right. 
This implies that $(a)$ contributes negatively to $A_N$, 
opposite to that of the associatively produced meson.  
Taking the contributions from the direct fusion processes 
(b) $u_v^P+(d_ss_s)^T\to \Lambda$ 
and (c) $d_v^P+(u_ss_s)^T\to \Lambda$ 
mentioned in section 4.2
and that from $N_0$ into account, 
we can obtain the $A_N$ for $\Lambda$ in different $x_F$ region. 
In fact, as shown in [\ref{BL96}], 
the above-mentioned differences and similarities 
between $A_N(x_F,\Lambda |s)$ 
and $A_N(x_F,\pi |s)$ directly reflect the interplay 
between the different direct fusion processes 
and the non-direct-formation part: 
In the region $x_F<0.4\sim 0.5$,  
(b), (c) and non-direct-formation part $N_0$ dominates. 
The situation is very much 
the same as that for $\pi^0$, hence we obtain 
that $A_N(x_F,\Lambda|s)$ in this region is similar to 
$A_N(x_F,\pi^0|s)$. 
For $x_F>0.4\sim 0.5$, (a) dominates thus 
$A_N(x_F,\Lambda|s)$ is similar to 
$-A_N(x_F,\pi^+|s)$. 

\begin{figure}
\begin{center}
\psfig{file=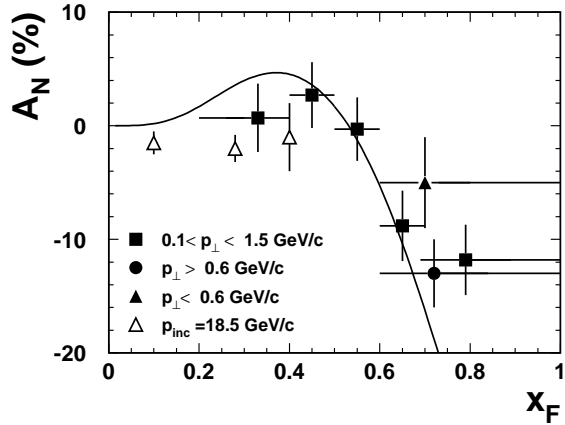,height=6cm}
\end{center}
\caption{Calculated $A_N$ as a function of $x_F$ for 
        $p(\uparrow)+p(0)\to \Lambda +X$. 
  Data are taken from [\protect\ref{E70495}]. 
This figure is taken from [\protect\ref{BL96}].}
\label{fig:ANLampp}
\end{figure}

The $x_F$-dependence of 
$A_N(x_F,\Lambda|s)$ has also been calculated\cite{BL96}.  
Now, the difference $\Delta N(x_F,\Lambda |s)\equiv 
N(x_F,\Lambda |s,\uparrow\nobreak) 
- N(x_F,\Lambda|s,\downarrow)$ 
contains contributions from (a), (b) and (c). 
The contributions from (b) and (c) are simply 
proportional to
$\Delta D_{b,c}(x_F,\Lambda |s)
\equiv D_{b,c}(x_F,\Lambda |s,+)-D_{b,c}(x_F,\Lambda |s,-)$; 
where  $D_{b,c}(x_F,\Lambda |s,\pm)$ are the number  densities 
for $\Lambda$'s formed the corresponding direct processes 
of upwards $(+)$ or downwards $(-)$ polarized valence quark 
[$u_v$ in case (b) and $d_v$ in case (c)] with suitable 
sea-diquarks. 
The contribution from (a) is opposite in sign to that 
of the associatively produced meson and is proportional 
to $- r(x|u_v,tr)\equiv -\Delta u_v(x|tr)/u_v(x)$, 
where $x$ is the fractional momentum of the $u_v$ valence quark. 
We have therefore, 
\begin{equation}
\Delta N(x_F,\Lambda |s)=C \Bigl [ 
-r(x|u_v,tr) D_a(x_F,\Lambda|s) + 
\Delta D_b(x_F,\Lambda|s) + 
\Delta D_c(x_F,\Lambda|s) \Bigr ], 
\label{eq:dNLam}
\end{equation} 
where $D_{a}(x_F,\Lambda |s)$ is given by Eq.(\ref{eq:DaLam}); 
$\Delta D_{b,c}(x_F,\Lambda |s,\pm )$ are given by, 
\begin{equation}
\Delta D_b(x_F,\Lambda|s) =\kappa _\Lambda  
      \Delta u_v(x^P|tr) f_D(x^T|d_ss_s),
\label{eq:dDblam}
\end{equation}
\begin{equation}
\Delta D_b(x_F,\Lambda|s) =\kappa _\Lambda  
       \Delta d_v(x^P|tr) f_D(x^T|u_ss_s).
\label{eq:dDclam}
\end{equation}
$A_N(x_F,\Lambda|s)$ can now be calculated 
using the ansatz for 
$q_v^\pm (x|tr)$ in Eqs.(\ref{eq:u+-u})-(\ref{eq:d+-d}). 
In this case $r(x|u_v,tr)=2/3$ is a constant. 
The results of this calculation 
are compared with the data\cite{E70495} 
in Fig.~\ref{fig:ANLampp}. 
We see that all the qualitative 
features of the data\cite{E70495} are well reproduced.

Similar analysis can be and have been made for 
other hyperons in a straight-forward way. 
Qualitative results have been obtained 
in [\ref{BL96}], 
which can be tested by future experiments.

\subsection{Further tests and developments}

At the end of this section, we would like to add 
a few comments concerning further tests 
and future developments and/or applications 
of this model. 

From the discussions in the last subsections, 
we see that the model has an advantage 
that the picture is very clear and the model is simple. 
In fact, in order to demonstrate the role 
played by the key factors, 
many other effects which are not essential in 
understanding the left-right asymmetries 
observed in single-spin reactions 
are simplified and/or neglected.
Semi-classical arguments are used in some places 
in order to make the picture as clear as possible. 
Thanks to this advantage, 
the model has a rather strong prediction power. 
In fact, as can be seen from the last few subsections, 
a number of predictions have 
been made in different connections. 
The model can therefore be tested easily. 
Till now, four of the predictions 
have already been confirmed by the experiments 
performed in the last years, many others 
can be tested by future experiments.
Here, we would like to add  
another testable characteristics of the model,  
i.e., according to the model,  
$A_N$'s in different reactions  
should have the following in common:

($\alpha $) $A_N$ for hadron production 
is expected to be significant in and only 
in the fragmentation region of the polarized colliding object.
It should be zero in the fragmentation region of the 
unpolarized one. 

($\beta $) $A_N$ for hadron production
in the fragmentation regions of the polarized colliding objects 
depends little on what kinds of unpolarized 
colliding objects are used.

($\gamma$) In contrast, $A_N$ for lepton-pair production 
depends not only on the polarized colliding object but also 
on the unpolarized one.

These common characteristics of $A_N$ for 
the production of different hadrons 
are consistent with the available data\cite{E70488,E70496b} 
and can be checked further 
by future experiments\cite{RHIC,HERAN}.

Despite of its successes in describing the available data, 
the model has its limitations. 
At the end, it is a phenomenological model where 
the physical picture is based on arguments some 
of which are semi-classical. 
A field theoretical description is still lacking.
This limits the prediction power and 
its applicability to other processes. 
In fact, there are many related questions 
which are still awaiting for answer.
E.g.: Can we calculate the constant $C$? Does it 
depends on transverse momentum $p_\perp$?
How is the energy dependence of $A_N$? 
How is the transverse momentum dependence of $A_N$?
These are questions which should be investigated. 
Furthermore, if it turns out that the model is right 
which implies that orbital motion of valence quarks and 
``surface effect'' are important in describing spin 
effects in hadron-hadron collisions, 
is it possible, if yes, how
to take them into account in the 
description of such reactions using 
perturbative QCD parton model?

\section{Testing different models by performing further experiments}

In the last two sections, we have discussed 
two different approaches. 
We have seen that both types of models have the potential
to give non-vanishing values for the left-right asymmetries 
in high energy single-spin hadron-hadron collisions.  
It is then natural to ask: 
``Can we tell which approach 
is the more appropriate one for the description of the 
above mentioned data$^{\ref{E70488}-\ref{E70496b}}$?'' 
The answer to this question is unfortunately ``No!''. 
The reason is that each of these models 
are based on a set of special assumptions. 
In particular, to obtain significant asymmetries 
in Type One models we have three possibilities, 
(i), (ii) and (iii);
but what has been measured$^{\ref{Kle76}-\ref{E70496b}}$ 
is the convolution of all three factors.
Hence it is difficult 
to find out which one is asymmetric 
simply through comparisons between 
the model and the existing data. 
Is it possible to have experiments 
with which the basic assumptions of these models can be tested 
{\em individually}\,?

This question has been discussed by different authors 
in different occasions (See, in particular, [\ref{BLMR96}] 
and the references given there.)
It has been shown that the question should at least partly 
be answered in the affirmative for  
the following reasons:

First, if we can determine the 
direction of motion of the polarized quark before 
it fragments into hadrons and 
measure the left-right asymmetry 
of the produced hadrons 
w.r.t. this (known as the jet-) direction, 
we can directly see whether the products 
of the quark hadronization process are
asymmetric w.r.t. this jet axis.
Such measurements should 
be able to tell us whether 
the corresponding quark fragmentation function is asymmetric. 

Second, there is no hadronization 
(i.e. no fragmentation) 
of the struck quarks in inclusive 
lepton-pair or $W^\pm$-production processes. 
Hence, measurements of the left-right asymmetries in 
such processes should yield useful
information on the properties of the factors other than the
quark-fragmentation function.   

Third, while surface effect 
and/or ``initial state interactions''
may play a significant role in 
hadron-hadron collision processes, 
they do not 
exist in deep inelastic lepton-hadron 
scattering in large $Q^2$ and 
large $x_B$ region where 
the exchanged virtual photons 
are considered as ``bare photons'' \cite{BLM96pro}. 
Hence, comparisons between these two 
processes can yield useful information on the 
role played by surface effects. 

Four experiments are listed in [\ref{BLMR96}].  
They are the following:

(A) Perform $l+p(\uparrow )\to l+\pi+X$ 
for large $x_B$ ($>0.1$, say) and large 
$Q^2$ ($>10$ GeV$^2$, say) 
and measure the left-right asymmetry 
in the current fragmentation region 
{\it w.r.t. the jet axis} (See, also, [\ref{Col93}]). 
Here, $l$ stands for charged lepton $e^-$ or  $\mu^-$; 
$x_B\equiv Q^2/(2P\cdot q)$ is the usual Bjorken-$x$, 
$Q^2\equiv -q^2$, and $P, k, k', q\equiv k-k'$ are the 
four momenta of the proton, incoming lepton, outgoing lepton 
and the exchanged virtual photon respectively. 
The $x_B$ and $Q^2$ are chosen in 
the abovementioned kinematic region 
in order to be sure that the following is true: 
(1) The exchanged virtual photon can be treated 
as a bare photon\cite{BLM96pro};  
(2) This bare photon will mainly 
be absorbed by a valence quark 
which has  significant polarization 
in a transversely polarized proton. 

In this reaction, 
a valence quark is knocked out by the virtual photon
$\gamma ^*$ and fragments into 
the hadrons observed in the current jet. 
The jet direction is approximately 
the moving direction of 
the struck quark before its hadronization; 
and the struck quark has a given probability to 
be polarized transversely to this jet axis 
The transverse momenta of the produced hadrons 
w.r.t. this axis
come solely from the fragmentation of the quark.   
Hence, by measuring this transverse momentum distribution, 
we can directly find out whether the fragmentation function 
of this polarized quark is asymmetric.

(B) Perform the same kind of experiments as 
that mentioned in (A) and  
measure the left-right asymmetry of the produced pions 
in the current fragmentation region 
{\it w.r.t. the photon direction} in the 
rest frame of the proton, and examine those events 
where the lepton plane is perpendicular to the 
polarization axis of the proton.
In such events, the obtained asymmetry should 
contain the contributions 
from the intrinsic transverse motion of quarks 
in the polarized proton and  
those from the fragmentation of polarized quarks, 
provided that they indeed exist. 
That is, we expect to  
see significant asymmetries, if and only if  
one of the abovementioned effects 
is indeed responsible for the asymmetries observed 
for pion production in single-spin 
hadron-hadron collisions. 

\begin{table}
\caption{Qualitative predictions for left-right asymmetries
in the discussed experiments
if the asymmetries observed in inclusive hadron production
in hadron-hadron collisions originate
from the different kinds of effects mentioned in the text.
See text for more details, in 
particular the signs and the magnitudes for $A_N$'s in the
cases where $A_N\not =0$. }
\label{tab:trac}
{\small
\begin{tabular}{c|c|c|c|c}   
\hline 
&
\multicolumn{4}{c}{
\begin{minipage}[t]{8.6cm}
\begin{center}  
\ \\[-0.3cm]
 If the $A_N$ observed in $p(\uparrow)+p(0)\to \pi +X$
 originates from ... 
\end{center}
\end{minipage}}\\[0.2cm]
\begin{minipage}[t]{5.23cm}
\begin{center}
  process \phantom{xxxx}
\end{center}
\end{minipage}
&
\begin{minipage}[t]{1.8cm}
\begin{center}
intrinsic quark distribution
\end{center}
\end{minipage}
&
\begin{minipage}[t]{1.8cm}
\begin{center}
elementary scattering process
\end{center}
\end{minipage}
&
\begin{minipage}[t]{1.8cm}
\begin{center}
quark fragmentation function
\end{center}
\end{minipage}
&
\begin{minipage}[t]{2.42cm}
\begin{center}
orbital motion of valence quarks \&
surface effect \\ \
\end{center}
\end{minipage}
\\[0.9cm]  \hline 
\begin{minipage}[t]{5.23cm}
\begin{center}
\ \\[-0.1cm]
 $l+p(\uparrow)\to l+
 \left(\begin{array}{cc}\pi^\pm\\ K^+\\ \end{array}\right) + X $
\end{center}
\end{minipage} &
\begin{minipage}[t]{1.8cm}
\begin{center}
\ \\[-0.3cm]
$A_N=0$ \\ wrt jet axis
\end{center}
\end{minipage}
&
\begin{minipage}[t]{1.8cm}
\begin{center}
\ \\[-0.3cm]
$A_N=0$ \\ wrt jet axis
\end{center}
\end{minipage}
&
\begin{minipage}[t]{1.8cm}
\begin{center}
\ \\[-0.3cm]
$A_N\not =0$ \\ wrt jet axis
\end{center}
\end{minipage}
&
\begin{minipage}[t]{1.8cm}
\begin{center}
\ \\[-0.3cm]
$A_N=0$ \\  wrt jet axis
\end{center}
\end{minipage}
\\ [-0.1cm] \cline{2-5}  
\begin{minipage}[t]{5.23cm}
\begin{center}
\ \\[-0.3cm]
 in the current fragmentation region \\
 for large $Q^2$ and large $x_B$
\end{center}
\end{minipage} &
\begin{minipage}[t]{1.8cm}
\begin{center}
\ \\[-0.3cm]
$A_N\neq 0$ \\ wrt $\gamma^\star$ axis
\end{center}
\end{minipage}
&
\begin{minipage}[t]{1.8cm}
\begin{center}
\ \\[-0.3cm]
$A_N=0$ \\ wrt $\gamma^\star$ axis
\end{center}
\end{minipage}
&
\begin{minipage}[t]{1.8cm}
\begin{center}
\ \\[-0.3cm]
$A_N\not =0$ \\ wrt $\gamma^\star$ axis
\end{center}
\end{minipage}
&
\begin{minipage}[t]{1.8cm}
\begin{center}
\ \\[-0.3cm]
$A_N=0$ \\ wrt $\gamma^\star$ axis
\end{center}
\end{minipage}
\\[0.6cm] \hline 
\begin{minipage}[t]{5.23cm}
\begin{center}
\ \\[-0.1cm]
 $l+ p(\uparrow) \to l+
 \left(\begin{array}{cc} \pi^\pm\\ K^+\\ \end{array}\right) + X $\\
 in the target fragmentation region \\
 for large $Q^2$ and large $x_B$
\end{center}
\end{minipage}
&
\begin{minipage}[t]{1.8cm}
\begin{center}
\ \\[0.4cm]  $A_N\neq 0$
\end{center}
\end{minipage}
&
\begin{minipage}[t]{1.8cm}
\begin{center}
\ \\[0.4cm] $A_N=0$
\end{center}
\end{minipage}
&
\begin{minipage}[t]{1.8cm}
\begin{center}
\ \\[0.4cm] $A_N\neq 0$
\end{center}
\end{minipage}
&
\begin{minipage}[t]{1.8cm}
\begin{center}
\ \\[0.4cm] $A_N=0$
\end{center}
\end{minipage}
\\[1.6cm] \hline 
\begin{minipage}[t]{5.23cm}
\begin{center}
\ \\[-0.1cm]
 $p + p(\uparrow)\to
\left(\begin{array}{cc} l\bar{l}\\ W^\pm\\ \end{array}\right)+ X $\\
 in the fragmentation region of $p(\uparrow)$
\end{center}
\end{minipage}
&
\begin{minipage}[t]{1.8cm}
\begin{center}
\ \\[0.2cm] $A_N\neq 0$
\end{center}
\end{minipage}
&
\begin{minipage}[t]{1.8cm}
\begin{center}
\ \\[0.2cm] $A_N\approx 0$
\end{center}
\end{minipage}
&
\begin{minipage}[t]{1.8cm}
\begin{center}
\ \\[0.2cm] $A_N=0$
\end{center}
\end{minipage}
&
\begin{minipage}[t]{1.8cm}
\begin{center}
\ \\[0.2cm] $A_N\neq 0$
\end{center}
\end{minipage}
\\ [1.3cm] \hline 
\end{tabular} }
\end{table}

(C) Perform the same kind of experiments as that in (A),
but measure the left-right asymmetry 
{\it in the target fragmentation region} 
w.r.t. the moving direction of the proton   
in the collider (e.g. HERA) laboratory frame. 
By doing so, we are looking at the fragmentation products 
of ``the rest of the proton'' complementary to the struck quark 
(from the proton). 
Since there is no 
contribution from the elementary 
hard scattering processes and there is no hadronic  
surface effect, $A_N$ 
should be zero if the existence of left-right asymmetries is due to 
such effects. 
But, if such asymmetries originate from the fragmentation  
and/or from the intrinsic transverse motion of the 
quarks in the polarized proton,  
we should also be able to see them here. 

(D) Measure the left-right asymmetry $A_N$ for $l\bar l$ and/or that for 
$W^\pm$ in $p(\uparrow )+p(0)\to l\bar l \ \mbox {or } W^\pm +X$. 
Here, if the observed $A_N$ for hadron production 
indeed originates from the quark fragmentation, we should see 
no left-right asymmetry in such processes. 
This is because there can be no  contribution 
from the quark fragmentation here.  
Hence, non-zero values for $A_N$ in such processes can only 
originate from asymmetric quark distributions --- 
including those due to  orbital motion of valence quarks 
and surface effect. 

The qualitative results are summarized in table \ref{tab:trac}. 
We emphasize here that the results of these experiments will 
not only shed light on the origin of the single-spin 
asymmetries for hadron production in hadron-hadron collisions but also 
yield extremely useful information on the widely used quark 
distribution functions 
in transversely polarized nucleon and the fragmentation functions. 
They should have strong impact on the study of spin 
distributions in nucleon and on the study of 
spin-dependent hadronic interactions.

\section{Orbiting valence quarks and hyperon polarization 
in inclusive production processes at high energies}

In a recent Letter\cite{LB97pol}, 
it has been pointed out
that there should 
be a close relation between 
the above discussed left-right asymmetries $A_N$
observed in single-spin hadron-hadron collisions 
and hyperon polarization $P_H$ observed$^{\ref{Les75}-\ref{WA8995}}$  
in unpolarized hadron-hadron collisions. 
There exist a large number of experimental indications 
and theoretical arguments which show that 
these two spin phenomena may have the same dynamical origin. 
We recall that the 
striking hyperon polarization in 
inclusive production processes 
at high energies has been discovered\cite{Les75,Bun76} 
already in 1970s.  
There has been a lot of interest  in studying 
the origin of this effect, 
both experimentally$^{\ref{Les75}-\ref{WA8995}}$ 
and theoretically$^{\ref{And79}-\ref{Sof92},\ref{LB97pol}}$.  
The pQCD parton model predicts  zero polarization\cite{Kane78}.    
It is now a well-known experimental 
fact$^{\ref{Les75}-\ref{WA8995}}$ 
that hyperons produced 
in high energy hadron-hadron or hadron-nucleus collisions
are polarized transversely to the production plane,  
although  the projectiles and  
the targets are unpolarized. 
Experimental results$^{\ref{Les75}-\ref{WA8995}}$  
have now been obtained for production of 
different  hyperons in reactions 
using different  projectile and/or 
different targets at different energies. 
But, theoretically, the origin still remains a puzzle. 
In fact, it has been considered as a standing challenge for the 
throretians to understand it for all these years. 
Hence, the result of [\ref{LB97pol}] is rather interesting 
since if it turns out to be true, 
it should certainly provide some clue in the  
searching of the origin of $P_H$.
We review the main ideas and results in the following.
For the sake of simplicity, we use the model for $A_N$ 
discussed in section 4 as an example.
We emphasize here that the discussion in [\ref{LB97pol}] 
depends in fact little on the model. 
We use this model in the following since we think 
such an example may be helpful 
in understanding the underlying physics 
of the discussions in [\ref{LB97pol}]. 
The readers are referred to the original paper for 
a more model independent discussion.

As has been mentioned above, 
there exist  a large amount of 
data$^{\ref{Les75}-\ref{WA8995}}$  
on hyperon polarization in unpolarized 
hadron-hadron collisions.
These data show  a number of striking characteristics 
very similar to  those of  $A_N$.  
In fact, we can simply replace $A_N$ by $P_H$ in 
(1) through (3) in section 2. 
These similarities already 
suggest that both phenomena have the same origin(s). 
Furthermore, we note: 
$A_N\not=0$ implies that  
the direction of transverse 
motion of the produced hadron depends on  
the polarization of the projectile. 
$P_H\not=0$ means that there exists a
correlation between the direction of transverse 
motion of the produced hyperon and 
the polarization of this hyperon. 
That is, both phenomena show the existence of 
correlation between transverse motion 
and transverse polarization.  
Hence, unless we (for some reason) 
insist on {\it assuming} that 
the polarization of 
the produced hyperons observed in 
the projectile-fragmentation region  
is independent of the projectile 
--- which would in particular contradict the empirical  
fact recently  observed  by E704 
Collaboration\cite{Bra95} for $\Lambda$ production --- 
we are practically forced to accept the conclusion that 
$A_N$ and $P_H$ are closely related to each other.  

This close relation has been studied in [\ref{LB97pol}] 
by considering the following questions: 
Can we understand the existence of $P_H$ and reproduce 
the main characteristics of the data if 
we use the data of $A_N$ as input?
Do we need further input(s)?
Two different cases, i.e.  
production of hyperon which has only one 
(or two) valence 
quark(s) in common with the projectile,   
have been considered separately in [\ref{LB97pol}]. 
In the following, we review these discussions 
using the model for $A_N$ as an example. 
  
In the model discussed in section 4,  
$A_N$ comes from the orbital motion of the valence quarks 
and the surface effect in hadron-hadron collisions. 
More precisely, the $A_N$ data, both those for meson 
and those for $\Lambda$ production, can be described 
using the following two points:

(I) Mesons $(M)$ and baryons $(B)$ produced through
$q_v^P+ \bar q_s^T \to M$  and
$q_v^P+ (q_sq_s)^T \to B$
have large probability to go left
(w.r.t. the collision axis
looking downstream)
if $q_v^P$ is upwards polarized
(w.r.t. the production plane).

(II) Baryons produced through
$(q_vq_v)^P+q_s^T \to B$ are
associated with $(q_v^a)^P+\bar q_s^T\to M$ and
have large probability to move in
the opposite transverse direction
w.r.t. the collision axis as $M$ does.

We note that these two points are direct implications 
of the picture in section 4. 
They can also be considered as direct implications\cite{Liang98} of 
the available $A_N$ data for mesons and $\Lambda$. 
We now use these two points as input to see 
if we can understand $P_H$. 
To do this, we encountered the following two questions: 

(1) Will the polarization of the quarks be retained in fragmentation? 

(2) What kind of picture for the spin structure of hadron 
shall we use in obtaining the polarization 
of the hadrons produced in fragmentation processes 
from the polarization of the quark(s) contained in the hadron: 
the one from the static quark model or that from 
polarized deep-inelastic 
lepton-nucleon scattering (DIS) data? 

These questions have been discussed in [\ref{BL98}].
We found out that $\Lambda$ is 
an ideal place to study these questions: 
First, the polarization of $\Lambda$ 
can easily be determined in 
experiments by measuring the angular distribution 
of its decay products. 
Second, $\Lambda$ has a very particular 
spin structure in the static quark model, i.e., 
$|\Lambda^\uparrow\rangle =(ud)_{0,0}s^\uparrow$ 
(here the subscripts of $ud$ denote its total spin and the third component),
where the spin of $\Lambda$ is completely carried 
by its $s$-valence quark. 
The results of the measurements\cite{Aleph96,OPAL} 
of longitudinal $\Lambda$ polarization 
in $e^+e^-\to Z\to \Lambda +X$ at LEP provided 
much insight into answering these questions. 
Since, according to the standard model for 
electroweak interactions, 
$s$-quarks from $Z$ decay are longitudinally polarized 
before hadronization. 
This longitudinal polarization 
can be transferred to the produced $\Lambda$, and 
the maximal value of longitudinal $\Lambda$ polarization 
is expected if the following two conditions are true: 

(1) Quark polarization is {\it not} destroyed in 
fragmentation.

(2) The SU(6) wavefunction 
from the static quark model is used to  
obtain the polarization of the produced hadron from 
that of the quark(s). 

This maximal expectation has been estimated in [\ref{GH93}]. 
The obtained result\cite{GH93} is indeed much larger 
than that obtained\cite{BL98} using the DIS picture, 
and is in good agreement with the ALEPH and OPAL data\cite{Aleph96,OPAL}. 
This strongly suggests that (1) and (2) are true.
 
\begin{figure}
\begin{center}
\psfig{file=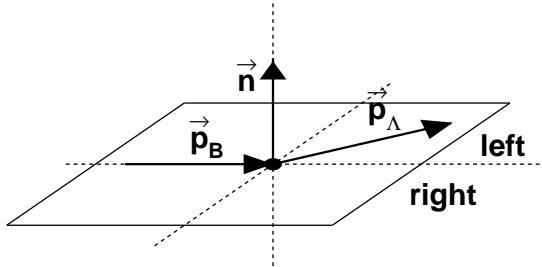,height=6cm}
\end{center}
\caption{Unit vector 
$\vec n\equiv 
\vec p_B\times \vec p_\Lambda/|\vec p_B\times \vec p_\Lambda|$,  
which is in the normal direction of the production plane and 
according to which hyperon polarization is defined. 
Here, $\vec p_B$ and $\vec p_\Lambda$ are momentum of the beam hadron 
and that of the produced $\Lambda$ respectively.  
We see in particular that $\vec n$ is pointing upwards if 
$\vec p_\Lambda$ is pointing to the left.}
\label{fig:koor}
\end{figure}

We now take points (I), (II) and (1) and (2) 
as inputs and check whether 
we can understand the $P_H$-data based on these points. 
We consider the first case discussed in [\ref{LB97pol}], 
i.e. the production of  hyperons 
which have only one valence quark in common with the projectile.   
In this case, hyperons in the fragmentation region are 
dominated by those containing this common valence quark 
from the projectile and a sea diquark from the target, 
i.e., they are mainly from the type of 
direct fusion process mentioned in (I). 
We ask the question whether we can understand 
$P_H$ in this case if we assume (I) is true. 
It can easily be shown that this question should be 
answered in the affirmative: 
$P_H$ in this case can be determined uniquely 
by (I) and the wave function of the hyperon. 
To see this, we recall that 
$P_H$ is defined w.r.t. the production plane.  
Hence, we need only to consider e.g. those hyperons 
which are going left and check 
whether they are upwards (or downwards) polarized.
(C.f. Fig.{\ref{fig:koor}.)
According to (I), if the hyperon is going left, 
$q_v^P$ should have a large probability to be upwards polarized.
This means, by choosing those hyperons which are going left, 
we obtain a subsample of hyperons which are formed by $q_v^P$'s 
that are upwards polarized with suitable sea diquarks.  
This information, together with 
the wave functions of the hyperons, 
determines whether the hyperon is polarized and, if yes, 
how large the polarization is. 
To demonstrate this, we consider 
$p+p\to \Sigma^-+X$. 
Here, the contributing direct fusion is 
$d_v^P+(d_ss_s)^T\to \Sigma^-$ and 
the wavefunction of $\Sigma^-$ is: 
$|\Sigma ^{-\uparrow} \rangle ={1\over 2\sqrt{3}}
     [3d^\uparrow (ds)_{0,0}+ d^\uparrow (ds)_{1,0}
      -\sqrt{2} d^\downarrow(ds)_{1,1} ]$,
where the subscripts of the diquarks are their total   
angular momenta and the third components.
We see that if $d_v^P$ is upwards polarized, 
$\Sigma^-$ has a probability of 
$2/3$ ($1/6$) to be upwards (downwards) polarized. 
Hence, we obtain that the $\Sigma^-$ from 
this direct formation process is 
positively polarized and the polarization 
is $(2/3)C$,  
[where $C$ is the positive constant mentioned 
in last section which describes 
the probability for $B$ from 
$q^P_v+(q_sq_s)^T\to B$ to 
go left if $q_v^P$ is upwards polarized.] 
Similar analysis can also be done 
for other hyperons and, e.g., 
we obtained that both $\Xi^-$ and $\Xi^0$ produced 
in $pp$-collisions are negatively polarized and 
the polarization is $-C/3$. 
These results show that 
$P_\Sigma^-$ 
is positive and its magnitude is large 
while $P_\Xi$ is negative and its magnitude is smaller. 
Measurements of both $P_\Sigma^-$ and $P_\Xi$  
in $pp$-collisions have been carried out$^{\ref{Les75}-\ref{Hel96}}$.
The results are consistent with the above mentioned expectations.

We see that these results 
follow directly from (I) together with (1) and (2) 
without any further input. 
There are also many other 
direct associations, e.g. the following: 

(A) $P_\Lambda$ in the beam fragmentation region 
    of $K^-+p\to \Lambda+X$ is large and is, in contrast to 
    that in $pp$-collisions, {\it positive} in sign.
This is because, according to the wave function, 
$|\Lambda^\uparrow \rangle =s^\uparrow (ud)_{0,0}$, 
the polarization of $\Lambda$ is entirely determined by the $s$ quark.
Here, the only contributing direct formation is 
$s^P_v+(u_sd_s)^T\to \Lambda$ and, according to (I), 
$s^P_v$ should have large probability to be upwards polarized 
if $\Lambda$ is going left.

(B) $P_\Lambda$ in $\pi^\pm+p\to \Lambda +X$ 
should be negative and the magnitude should be very 
small in the $\pi$ fragmentation region. 
This is because, in this region, 
the only contributing 
direct formation process 
is $u_v^P+(d_ss_s)^T\to \Lambda$ 
[or $d_v^P+(u_ss_s)^T\to \Lambda$] 
and the $\Lambda $ directly produced 
in this process is unpolarized. 
A small polarization is expected 
only from the decay of $\Sigma^0$.

(C) Not only the produced hyperons 
but also the produced vector mesons 
are expected to be transversely polarized
in the projectile fragmentation region 
of hadron-hadron or hadron-nucleus collisions. 
E.g., vector mesons such as $\rho ^\pm, \rho^0, K^{*+}$ 
produced in $p+p$ collisions 
are expected to be positively polarized 
in the proton  fragmentation region. 
This is because, if such mesons are going left, 
the valence quarks which combine 
with suitable anti-sea-quarks 
to form such mesons should 
have large probability to be upwards polarized. 
This upwards polarized valence quark combine with 
a suitable (unpolarized) anti-sea-quark and 
leads to a vector meson which has 
large probability to be upwards polarized.

(D) There should be no significant polarization 
transverse to the production plane 
 for hyperons produced 
in $e^+e^-$-annihilations. 
This is because, in such processes there is  
neither hadronic surface effect 
nor orbital motion of the initial state quarks.  

(E) There should be no significant polarization 
transverse to the production plane 
for hyperons produced 
in deep inelastic lepton-hadron collision process 
in the large $Q^2$ and large $x_B$ region. 
This is because, in this kinematic region, 
the exchanged virtual photons are ``bare photons''. 
They are  point-like objects and  
hence  there is no hadronic  surface effect 
in the photon-proton reactions.

Such direct associations can be 
readily checked by experiments. 
There are already  data 
for the processes mentioned 
in (A), (B) and (D) $^{\ref{Ben83}-\ref{WA8995}}$, 
and all of them are {\it in agreement with 
these associations}. 
More precisely, we see the following:  
Experiments using $K^-$ beams have been carried out and 
the results\cite{Ben83} show that  $P_\Lambda$ in the 
beam fragmentation region is indeed positive and large.
There are also experiments using $\pi$ beam and the results
show that $P_\Lambda$ in beam fragmentation is indeed much 
smaller than that in $p+p\to \Lambda +X$.
$\Lambda$-polarization in processes $e^+e^-\to $ hadrons 
has also been studied by 
TASSO Collaboration at PETRA DESY\cite{Tasso85} 
and ALEPH at LEP CERN\cite{Aleph96}. 
The data from both Collaborations 
show that $P_\Lambda$ 
is much less significant than that 
in $p+A$-collisions. 
This is consistent with (A).  
(C) and (E) can be checked by future 
experiments.\footnote{We note here, since vector mesons such as $\rho$ 
and $K^*$ decay into two hadrons via strong interactions, 
it is impossible to determine whether they are upwards or downwards 
polarized w.r.t. the production plane by measuring these decay products. 
This means to test (C) is rather academic according to the present technic 
of measuring the polarization of the produced particles. 
However, one can determine whether they are transversely or 
longitudinally polarized by measuring the angular distributions 
of these decay products. 
[See, K. Schiling and G. Wolf, Nucl. Phys. B{\bf 61}, 381 (1973).]
This means that (C) can at least be tested 
partly using the technology we know presently.}

These results are rather encouraging. 
Hence, we continue  
to consider the second case, 
i.e., the production of hyperon which has two valence quarks 
in common with the projectile. 
In this case hyperon containing 
a valence diquark of the projectile 
with a suitable seaquark of the target 
dominates the beam fragmentation region. 
The most well known process of this type is 
$p+p\to \Lambda +X$. In this process, the  
dominating contribution in the fragmentation is from 
the direct fusion process 
(a) $(u_vd_v)^P+s_s^T\to \Lambda $
mentioned in sections 4.2 and 4.5.
This direct fusion process (a) 
is mainly associated with 
$(u_v^a)^P+\bar s_s^T\to K^+$; and, 
according to (II),
if the $\Lambda$ is going left,  
the associatively produced $K^+$ 
should have a large probability to go right. 
This implies that $(u_v^a)^P$ has a large
probability to be downwards polarized. 
Since $K$ is a pseudoscalar meson and 
thus a spin-zero object, 
$\bar s_s^T$ should be upwards polarized. 
Hence, the corresponding $s_s^T$ should be downwards polarized,  
{\it provided that the sea quark-anti-quark pair is 
not transversely polarized}. 

We see that, to explain the existence 
of $P_\Lambda$ in this case,
we need, besides (II), a further assumption, i.e.,  
the sea quark-antiquark pair $s\bar s$ is not transversely polarized. 
As has been mentioned in [\ref{LB97pol}] that 
this can be considered as a further implication of the existence 
of $P_\Lambda$ for the structure of nucleon in the framework 
of the picture described in section 4. 
Whether it is indeed true can and should be checked by 
further experiments. 
It is therefore important to consider the 
direct consequences of this assumption  
and compare them with the available data.
For this reason, the following 
have been made in [\ref{LB97pol}].

First, a similar analysis has been made for the production 
of other hyperons of this type. 
Qualitative results for their $P_H$'s 
have been obtained and    
they are all consistent with the available data\cite{Hel96}. 

Second, a quantitative estimation 
of $P_\Lambda$ in $p+p\to\Lambda+X$ 
as a function of $x_F$ has been made. 
Since all the different  contributions 
[i.e. those from the direct fusion (a), (b) or (c), or 
the non-direct formation part $N_0$] and also the constant $C$ 
are known (see sections 4.2 and 4.5), 
this estimation can be made without 
any free parameter.
The results\cite{LB97pol} 
are compared with the data$^{\ref{Smi87}-\ref{Ram94}}$
in Fig.\ref{fig:Lampol}.\footnote{We note that, 
in this estimation, 
only associated production of $\Lambda$ 
and $K^+$ is considered 
while that of $\Lambda$ and $K^{*+}$ is neglected. 
It is clear that 
the above mentioned $\Lambda$ polarization 
can be obtained in the former case 
but completely destroyed in the latter. 
It is also clear that associated production of
$\Lambda$ and $K^{*+}$ contributes also significantly 
to $pp\to \Lambda X$. 
Inclusion of this contribution should reduce 
$|P_\Lambda|$ in the large $x_F$ region.
In this sense, the curve in Fig.13 
in the large $x_F$ region represents only 
the maximal expectation  of $|P_\Lambda|$ from the model. 
From the figure, we see also that 
$|P_\Lambda|$ in this region is 
indeed higher than the data, 
which implies that there is room for including such influences.
However, a detailed calculation in which all  effects 
such as associated production of $\Lambda$ and $K^*$ 
are taken into account needs a rather detailed 
hadronization model which describes  
the production of different  hadrons 
in the fragmentation region.
Although there exist several fragmentation models on the market,  
it is unclear whether they can  
describe vector meson production, 
in particular vector to pseudoscalar ratio,
in the fragmentation regions since no such study has been made yet.
This implies that such a detailed calculation would involve 
rather high theoretical uncertainties. 
Hence, we choose not to do such a calculation 
but seek for further direct tests of the picture. 
See in particular point ($\gamma$) in the following and  
our discussion of $P_\Lambda$ in 
$pp\to \Lambda K^+p$ at the end of this section.}

\begin{figure}
\begin{center}
\psfig{file=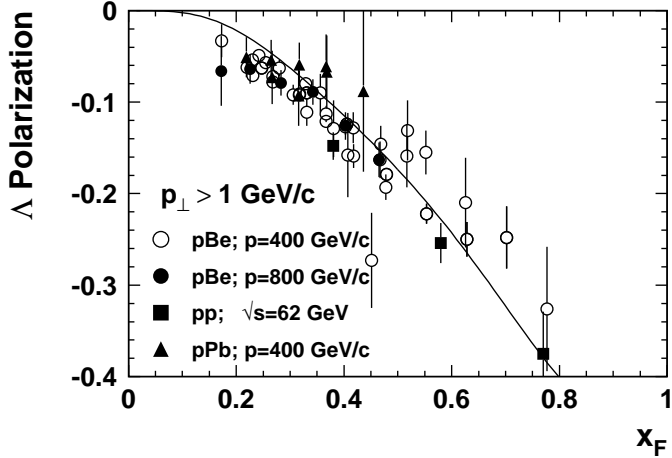,height=6.6cm}
\end{center}
\caption{Calculated results  
for $P_\Lambda$ as a function of $x_F$.
Data are taken from [\protect\ref{Smi87}-\protect\ref{Ram94}].
This figure is taken from [\protect\ref{LB97pol}].} 
\label{fig:Lampol}
\end{figure}

Third, a number of 
other consequences have been derived. 
The following are three examples which are 
closely related to the assumption that the 
$s$ and $\bar s$ which take part in the associated production 
are opposite in transverse spins.

($\alpha$) The polarization of the projectile 
and that of $\Lambda$ in the fragmentation region 
of $p+p\to \Lambda +X$ should be closely 
related to each other. 
In other words, the spin transfer 
$D_{NN}$ (which is defined as the probability 
for the produced $\Lambda$ to be upwards polarized 
in the case that the projectile proton is upwards polarized)
is expected to be positive 
and large for large $x_F$.   
It is true that the $ud$-diquark 
which forms the $\Lambda$ is in 
a spin-zero state thus carries no 
information of polarization.  
But, according to the mechanism of 
associated production, 
the polarization of the left-over $u_v^P$ 
determines the polarization of the projectile and that 
of the $s_s$ quark which combines with the 
$ud$-diquark to form the $\Lambda$. 
Hence, we expect to see 
a strong correlation between the polarization of 
the proton and that of the $\Lambda$.
A quantitative estimation of $D_{NN}$ as a 
function of $x_F$ is made.  
The result is shown in Fig.\ref{fig:DNN}.

\begin{figure}
\begin{center}
\psfig{file=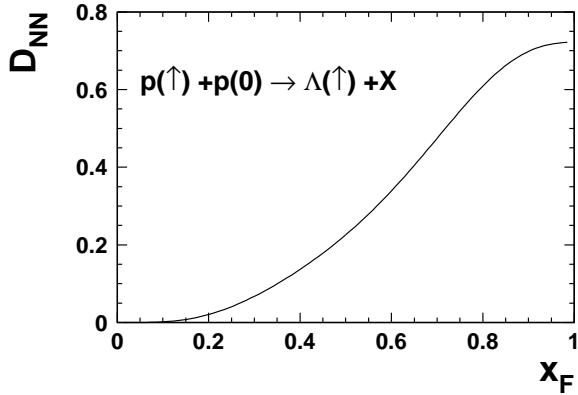,height=5.6cm}
\end{center}
\caption{$D_{NN}$ as a function of $x_F$ 
calculated by using  the proposed model  
for the case that the correlation between the 
spin of the $s$-sea-quark  [which forms together with the 
$(ud)_{00}$ valence diquark the $\Lambda$]
and the spin of the remaining $u$-valence quark in the projectile 
(which forms together with the $\bar s$-sea-quark 
the associated $K^+$) is maximal.
In this sense, it stands for the 
upper limit of our expectation. 
This figure is taken from [\protect\ref{LB97pol}].} 
\label{fig:DNN}
\end{figure}

($\beta$) $P_\Lambda$ in the beam fragmentation region 
of $\Sigma^-+A\to \Lambda +X$ 
should be {\it negative} and much less significant 
than that in $p+p\to\Lambda+X$.  
Here, the dominating contributions are 
the $\Lambda$'s which consist of 
$(d_vs_v)^P$ and $u_s^T$, 
$d_v^P$ and $(u_ss_s)^T$,  
or $s_v^P$ and $(u_sd_s)^T$.
Exactly the same analysis as that mentioned 
above for $p+p\to\Lambda+X$ show that
the $\Lambda$'s of the first two kinds are unpolarized; 
and those of the third kind   
are positively polarized. 
Since the first kind dominates in the 
large $x_F$ ($x_F>0.6$, say, this should be the 
same as that for $p+p\to \Lambda+X$, see section 4.2), 
the second and the third 
dominate the middle $x_F$ ($x_F\sim 0.3-0.4$) region,  
we expect to see the following:
If we exclude the contribution from 
$\Sigma^0$ and $\Sigma^{*0}$ decay, 
$P_\Lambda$ should be approximately zero for 
large $x_F$ and should be small but 
positive in the middle $x_F$ region. 
Taking $\Sigma^0$ and $\Sigma^*$ decay into account, 
we expect a small negative $P_\Lambda$ for large $x_F$. 

($\gamma$) Hyperon polarization in  
processes in which a vector meson is associatively produced 
should be very much different from that in  
processes in which a pseudoscalar 
meson is associatively produced.  
E.g., $P_\Lambda$ in the fragmentation region of   
$p+p\to \Lambda +K^{+}+X$ should be negative 
and its magnitude should be large, but  
$P_\Lambda$ in the fragmentation region of  
$p+p\to \Lambda +K^{*+}+X$ should be positive 
and its magnitude should be much smaller.
This is because, in the latter case, 
using the same arguments as we used in the former case,
we still obtain that $(u_v^a)^P$ 
(contained in $K^{*+}$) 
has a large probability to be downwards polarized 
if $\Lambda $ is going left.  
But, in contrast to the former case, 
the $\bar s_s^T$ here in the $K^{*+}$ can be upwards 
or downwards polarized since $K^{*+}$
is a spin-1 object. 
If $\bar s_s^T$ is upwards polarized, 
the produced meson can either be a $K^*$ or a $K$, 
and the corresponding $\Lambda$ should be 
downwards polarized. 
But if $\bar s_s^T$ is downwards polarized, 
the produced meson can only be a $K^{*+}$ and 
the corresponding $\Lambda$ should 
be upwards polarized, i.e. $P_\Lambda>0$.

Presently, there are data available 
for the processes mentioned 
in ($\alpha$) and ($\beta$) \cite{Bra95,WA8995}, 
and both of them are in agreement with 
the above expectations. 
More precisely, we have the following:  
E704 Collaboration has recently observed\cite{Bra95} 
a strong correlation between the polarization 
of the proton projectile and that for $\Lambda$ 
in the fragmentation region. 
This seems, at the first sight, difficult 
to be understood since the $ud$-diquark 
which is common for $\Lambda$ and proton 
should be in the spin zero state thus carries 
no information of polarization, but is    
a natural consequence [see ($\alpha$)] 
of the proposed picture. 
WA89 Collaboration has recently 
found out\cite{WA8995} that $P_\Lambda$
in the beam fragmentation region 
in $\Sigma^-+A\to \Lambda+X$ 
is negative in sign but indeed much less 
significant than that in $p+A\to \Lambda+X$,   
which is consistent with ($\beta$).
The prediction mentioned in ($\gamma$) 
is another characteristic property of the model 
and can be used as a crisp test of the picture.

We note that\cite{LB98Diff} 
a characteristic property of the proposed 
picture is that $\Lambda$ polarization 
in $pp\to \Lambda X$ comes predominately 
from those $\Lambda$'s each of which contains a 
valence diquark $(ud)_{0,0}$ of the colliding proton 
and is associated with a spin-zero meson such as $K^+$ 
which contains the other valence $u_v$ of that proton. 
Hence, if these conditions are guaranteed 
in a particular channel for $pp\to \Lambda X$, 
$|P_\Lambda|$ should take its maximum in this channel 
and the maximal value is $C$.
There exists indeed such a channel, i.e.  
$pp\to \Lambda K^+p$. 
In fact this is the only channel 
in which the above-mentioned 
two conditions are guaranteed. 
$\Lambda$ polarization in $pp\to \Lambda K^+p$ 
has been studied\cite{R608} by R608 Collaboration at CERN.
They obtained that $P_\Lambda =-0.62\pm0.04$. 
This is in excellent agreement with the above expectation 
$|P_\Lambda|=C=0.6$. 
This is strong support of the picture.

\section{Probing dissociation of space-like photons 
in deep-inelastic lepton-nucleon scattering} 

As have been seen in the previous sections, 
although the origin is still in debate, 
it is now a well know fact that striking left-right 
asymmetry $A_N$ exists in single-spin hadron-hadron collisions 
and striking transverse polarization $P_H$ exists 
for hyperon production in unpolarized 
hadron-hadron collisions. 
It is therefore natural to think about  
using such striking spin effects as a tool to study 
the properties of the particles and/or of strong interactions. 
A well-known example for the experimentalists is the 
use of the existence of $A_N$ as a polarimeter which has 
been discussed in different occasions. 
Another example has been discussed in [\ref{BLM96pro}]. 
We make a brief introduction of the latter here.

It is known already for a long time that 
hadronic dissociation of space-like
photons may play a
significant role in deep-inelastic 
lepton-hadron scattering --- especially
in diffractive processes\cite{Nieh70,Bauer78}. 
People seem to agree  
(for a list of references, see [\ref{BLM96pro}])
that, viewed from the hadron- or nucleus-target, 
not only real, but also space-like photons 
$(Q^2\equiv -q^2>0$, 
where $q$ is the four-momentum of such a photon) 
may exhibit hadronic dissociation. 
In the small $x_B$ region such as that at HERA, 
the coherent length of such hadronic dissociation 
may be as long as $10^3$ fm, much much longer than the 
typical size of a hadron. 
What does this mean for the interaction of such 
virtual photon $\gamma^*$ with hadrons? 
Does this imply that, viewed from the hadron, 
such photon behaves always like a hadron 
so that we have always a hadron-hadron collision in 
inelastic lepton-nucleon scattering in the 
small $x_B$ region?

In the above sections, we summarized that a characteristic 
property of hadron-hadron collisions is the existence 
of left-right asymmetry $A_N$ in the fragmentation region 
in single-spin processes 
and the existence  of hyperon polarization 
in the fragmentation 
of unpolarized collision processes. 
Although the origins of such striking 
spin effects are still in debate, 
both experimental results and theoretical arguments 
seem to suggest the following: 
there is no such spin effects in processes where 
the hadronic surface effect or 
``initial state interaction'' does not exist. 
Hadronic surface or initial state interaction 
exists in any hadron-hadron collision 
but should not exist in $\gamma^*$-hadron 
collision if $\gamma^*$ can be treated as pointlike. 
Hence, we can use the existence of such spin effects 
as a sensor to test whether the virtual photon 
behaves like a hadron. 

To be more precise, 
it has been proposed\cite{BLM96pro} 
that one can perform single-particle 
inclusive measurements in the proton fragmentation region
at HERA for small-$x_B$ and different $Q^2$ 
in reactions using transversely polarized or 
unpolarized protons, and compare the 
results with those 
in the corresponding hadron-hadron collisions.
Two extreme cases have been discussed\cite{BLM96pro}. 
In case one, $\gamma ^*$ is assumed to behave always like 
a hadron in the same $x_B$ region independent of $Q^2$. 
In this case, one expects to see $A_N$ for $\pi$'s or $K$'s 
in reactions using transversely polarized protons, 
or transverse polarization  $P_\Lambda$ 
for $\Lambda$ 
in reactions using unpolarized protons.  
The $A_N$'s and $P_\Lambda$ should be the same as those 
observed in the corresponding hadron-hadron collisions 
and they should be independent of $Q^2$. 
In the second case, the virtual photon $\gamma^*$ 
is assumed to behave like a hadron only in the small 
$Q^2$ region, where vector meson dominance plays a role,  
 but behaves like a pointlike object for large $Q^2$ 
independent of $x_B$.
If this is true, one should see a strong $Q^2$-dependence 
for $A_N$ and $P_\Lambda$. 
They should be approximately the same as those in the 
corresponding hadron-hadron collisions if $Q^2$ is small, 
but tends to zero for large $Q^2$. 
This has been demonstrated\cite{BLM96pro} 
in a quantitative manner by 
separating $F_2^p(x_B,Q^2)$-data\cite{NMC92,E66594} 
in the small-$x_B$ region into  
the well-known vector-dominance contribution 
(See e.g. [\ref{Pil95}] and the references cited there) 
from ``the rest'' which should be identified as 
the contribution from ``pointlike'' photon in this case. 
This separation is shown in Fig.\ref{fig:F2sep}.
In this way, the $Q^2$ dependence of $A_N$ and that 
of $P_\Lambda$ have been obtained and are shown in 
Figs.\ref{fig:ANphoton} and \ref{fig:PHphoton}.

\begin{figure}
\begin{center}
\psfig{file=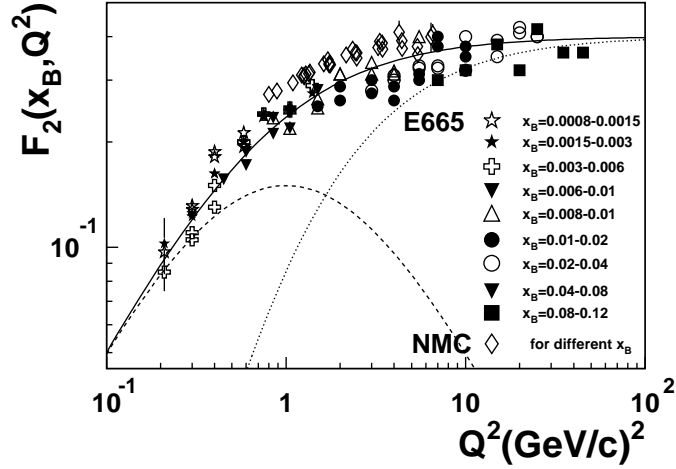,height=6.6cm}
\end{center}
\caption{Structure function $F_2^p(x_B,Q^2)$ as a function of $Q^2$. 
The data-points are taken from [\protect\ref{NMC92}] and 
[\protect\ref{E66594}]; 
and they are parameterized by the solid line. 
The dashed line is the contribution from the 
vector meson dominance. 
The dotted line is ``the rest''. 
The figure is taken from [\protect\ref{BLM96pro}]. }
\label{fig:F2sep} 
\end{figure}

\begin{figure}
\begin{center}
\psfig{file=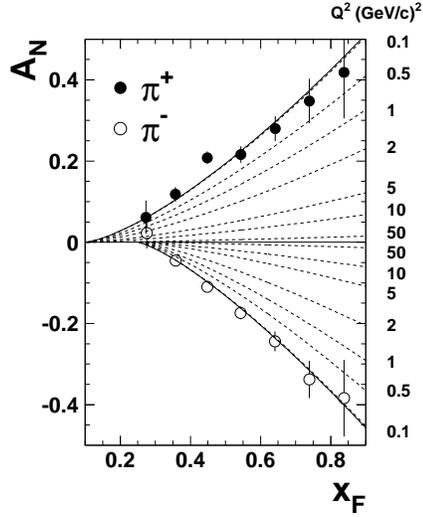,height=6.6cm}
\end{center}
\caption{$Q^2$ dependence of $A_N$ in 
$p(\uparrow ) +\gamma^*  \rightarrow \pi^\pm +X$ as a function of 
$x_F$ in case two discussed in the text. 
The data are for $p(\uparrow ) +p \to \pi^\pm +X$ 
and are the same as those in Fig.1. 
This figure is taken from [\protect\ref{BLM96pro}].}
\label{fig:ANphoton}
\end{figure}

\begin{figure}
\begin{center}
\psfig{file=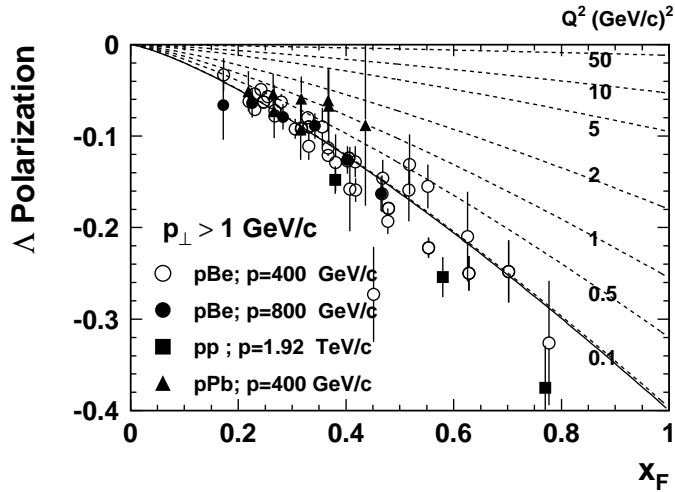,height=6.6cm}
\end{center}
\caption{$Q^2$ dependence of $P_\Lambda$ in 
$p +\gamma^*  \rightarrow \Lambda  +X$ as a function of 
$x_F$ in case two discussed in the text. 
The data are taken from [\protect\ref{Smi87}-\protect\ref{Ram94}]. 
The figure is taken from [\protect\ref{BLM96pro}]. }
\label{fig:PHphoton}
\end{figure}

We see from the figures that the differences between 
the results obtained in 
the two cases are significantly large. 
We therefore expect that they can indeed be used as a 
good probe for the hadronic dissociation of the photon.
From this example, we also explicitly see the following: 
Single-spin asymmetry study is interesting 
not only because the understanding of its origin 
can provide us a great deal of useful information 
on the structure of hadron and on the properties 
of hadronic interactions but also 
because it can be used as an useful tool to study 
the properties of different elementary particles and 
those of high energy reactions in general.

Acknowledgment

It is a great pleasure for both of us to express our 
sincere thanks to our common supervisor, 
Professor Meng Ta-chung, 
for bringing us to this interesting field, 
for the continuous guidance, 
the inspiring discussions and encouragements 
at every step of our works. 
We thank also R. Rittel for collaborations and discussions. 
Our sincere thank also goes to Professors M. Anselmino, 
D.H.E. Gro\ss{}, K. Heller, S.~Nurushev, J. Qiu and A. Yokosawa 
for helpful discussions in different occasions.   
This work was supported in part by 
National Natural Science Foundation of China (NSFC), 
by the State Education Commission of China, 
by the Australian Research Council and by 
Deutsche Forschungsgemeinschaft (DFG Me470-2).

\end{document}